\pdfoutput=1

\documentclass[11pt,twoside,a4paper,cmspaper,final,collab]{cms-tdr}
\def\svnVersion{452291P}\def\cmsCernNoTag{CERN-EP-2017-141}\def\cmsCernDate{\today}\def\cmsMessage{Published in the European Physical Journal C as \href{http://dx.doi.org/10.1140/epjc/s10052-018-5691-6}{\doi{10.1140/epjc/s10052-018-5691-6.}}}

\begin{document}\cmsNoteHeader{FSQ-12-001}

\hyphenation{had-ron-i-za-tion}
\hyphenation{cal-or-i-me-ter}
\hyphenation{de-vices}
\RCS$Revision: 452137 $
\RCS$HeadURL: svn+ssh://svn.cern.ch/reps/tdr2/papers/FSQ-12-001/trunk/FSQ-12-001.tex $
\RCS$Id: FSQ-12-001.tex 452137 2018-03-22 06:09:06Z robertc $
\newlength\cmsFigWidth
\ifthenelse{\boolean{cms@external}}{\setlength\cmsFigWidth{0.98\columnwidth}}{\setlength\cmsFigWidth{0.75\textwidth}}
\newlength\cmsTabSkip\setlength\cmsTabSkip{1.5ex}
\newlength\ruleht\setlength{\ruleht}{\baselineskip}
\providecommand{\pmrule}{\rule[-0.35\ruleht]{0pt}{1.15\ruleht}}

\ifthenelse{\boolean{cms@external}}{\providecommand{\cmsLeft}{top}}{\providecommand{\cmsLeft}{left}}
\ifthenelse{\boolean{cms@external}}{\providecommand{\cmsRight}{bottom}}{\providecommand{\cmsRight}{right}}
\cmsNoteHeader{FSQ-12-001}
\title{Study of dijet events with a large rapidity gap between the two leading jets in pp collisions at $\sqrt{s}=7$\TeV}

\date{\today}

\abstract{
Events with no charged particles produced between the two leading jets are studied in proton-proton collisions at $\sqrt{s}=7$\TeV. The jets were required to have transverse momentum $\pt^{\text{jet}}>40$\GeV and pseudorapidity $1.5<\abs{\eta^{\text{jet}}}<4.7$, and to have values of $\eta^{\text{jet}}$ with opposite signs. The data used for this study were collected with the CMS detector during low-luminosity running at the LHC, and correspond to an integrated luminosity of 8\pbinv. Events with no charged particles with $\pt>0.2$\GeV in the interval $-1<\eta < 1$ between the jets are observed in excess of calculations that assume no color-singlet exchange. The fraction of events with such a rapidity gap, amounting to 0.5--1\% of the selected dijet sample, is measured as a function of the \pt of the second-leading jet and of the rapidity separation between the jets. The data are compared to previous measurements at the Tevatron, and to perturbative quantum chromodynamics calculations  based on the Balitsky--Fadin--Kuraev--Lipatov evolution equations, including different models of the non-perturbative gap survival probability.
\\[3ex]
\textit{We dedicate this paper to the memory of our colleague and friend Sasha Proskuryakov, who started this analysis but passed away before it was completed. His contribution to the study of diffractive processes at CMS is invaluable.}
}

\hypersetup{%
pdfauthor={CMS Collaboration},%
pdftitle={Study of dijet events with a large rapidity gap between the two leading jets in pp collisions at 7 TeV},%
pdfsubject={CMS},%
pdfkeywords={CMS, physics, QCD, diffraction}}

\maketitle
\section{Introduction}

In high-energy proton-proton collisions, an interaction with large momentum transfer between two partons may lead to the production of a pair of jets with large transverse momenta \pt.
Dijet production at the LHC~\cite{Khachatryan:2011zj,Chatrchyan:2011qta,Chatrchyan:2012gwa,Chatrchyan:2012pb,Khachatryan:2016hkr,Sirunyan:2017skj,Aad:2010ad,Aad:2011jz,Aad:2011fc,Aad:2013tea,Aad:2014pua,Aad:2015xis} is generally well described by perturbative quantum chromodynamics (pQCD) calculations based on the Dokshitzer--Gribov--Lipatov--Altarelli--Parisi (DGLAP) evolution equations~\cite{dglap1,dglap2,dglap3}. The DGLAP equations govern the emission of additional softer partons, ordered in transverse momentum $k_\mathrm{T}$ with respect to the jets axes. However, when the two jets are separated by a large interval in pseudorapidity ($\eta$), an alternative pQCD evolution based on the Balitsky--Fadin--Kuraev--Lipatov (BFKL) equations~\cite{bfkl1,bfkl2,bfkl3} is expected to describe the data better~\cite{bj}. In the BFKL approach, the emission of additional partons is ordered in $\eta\sim\ln(1/x)$, where $x$ is the fractional momentum carried by the radiated parton.

\begin{figure}
\centering
\includegraphics[width=\cmsFigWidth]{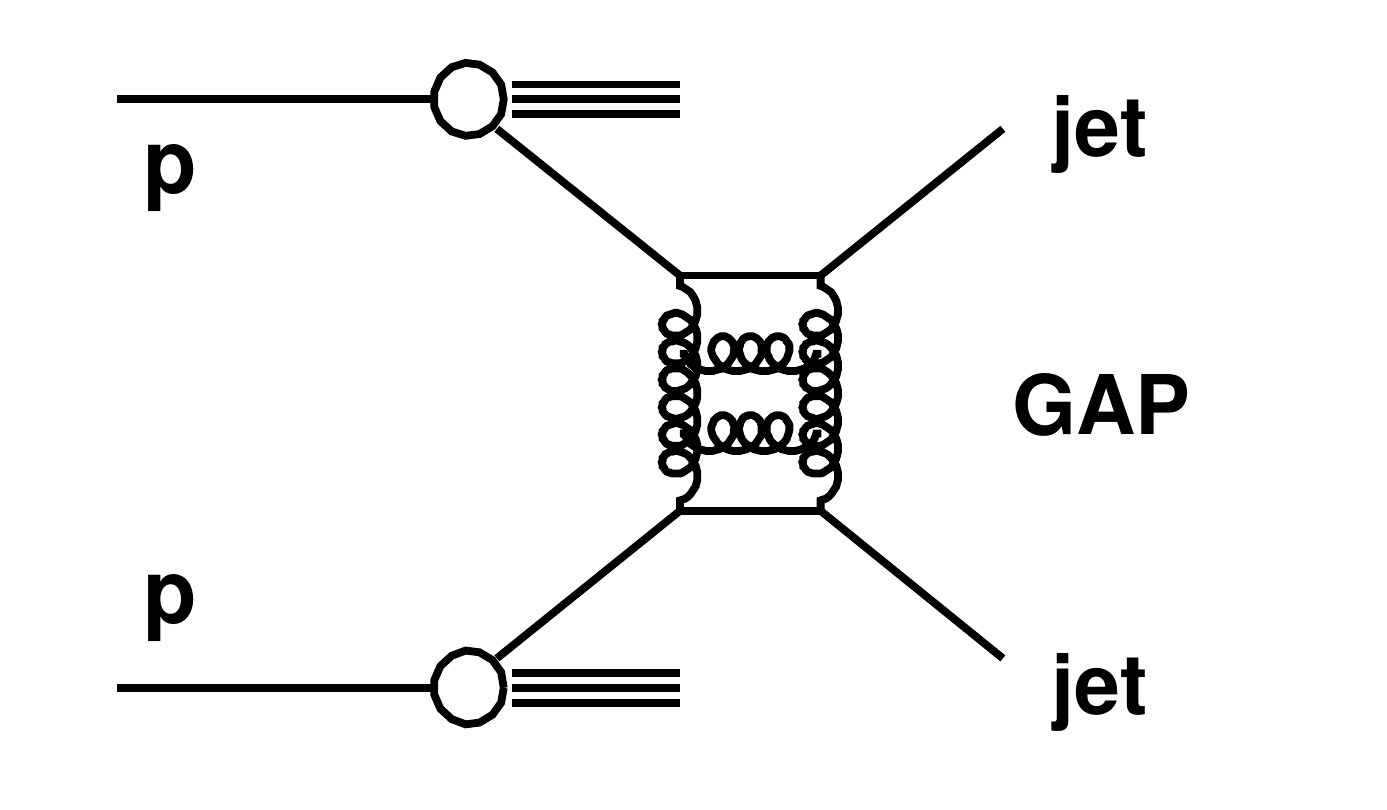}
\caption{Schematic diagram of a dijet event with a rapidity gap between the jets (jet-gap-jet event). The gap is defined as the absence of charged particle tracks above a certain \pt threshold.}
\label{fig-jgj}
\end{figure}

The events considered in this study are pp collisions where two jets are produced with a large rapidity gap between them. The absence of particles between the jets is reminiscent of a diffractive process~\cite{diffraction}, in which a color-singlet exchange (CSE) takes place between the interacting partons. In diffractive processes, such an exchange is described in terms of the pomeron, a combination of gluons in a color-singlet state. However, the absolute value of the four--momentum squared exchanged in standard diffractive events (less than a few $\GeV^2$) is much smaller than that in the events considered here. Such events can be understood in a BFKL-inspired approach in terms of the exchange of a color-singlet gluon ladder (Fig.~\ref{fig-jgj}), as first discussed by Mueller and Tang in Ref.~\cite{mt} and further developed in Refs.~\cite{csp,cspLHC,kmr}. Jet-gap-jet events in proton--proton collisions may be affected by additional scatterings among the spectator partons, which can destroy the original rapidity gap. Such a contribution is typically described by a non-perturbative quantity, the so-called gap survival probability, which quantifies the fraction of events where the rapidity gap is not destroyed by interactions between spectator partons~\cite{bj}.

Jet-gap-jet events were first observed in $\Pp\PAp$ collisions at the Tevatron by D0~\cite{d01,d02,d03} and CDF~\cite{cdf1,cdf2,cdf3}, and in $\Pepm\Pp$ collisions at HERA~\cite{zeus,h1}. At the Tevatron, the fraction of dijet events produced through CSE was found to be $\sim$1\% at $\sqrt{s}=1.8$\TeV,
a factor of 2--3 less than at $\sqrt{s}=0.63$\TeV. This paper presents the first observation of jet-gap-jet events at the LHC, and the measurement of the CSE fraction at $\sqrt{s}=7$\TeV, using events with two leading jets of $\pt^{\text{jet}}>40$\GeV and $1.5<\abs{\eta^{\text{jet}}}<4.7$, reconstructed in opposite ends of the CMS detector. The CSE signal is extracted from the distribution of the charged-particle multiplicity in the central region $\abs{\eta}<1$ between the jets, for particles with $\pt>0.2$\GeV. The CSE fraction is studied as a function of the pseudorapidity separation $\Delta\eta_\mathrm{jj}$ between the jets, and of the \pt of the second-leading jet, as done by the D0 experiment~\cite{d03}.

The data used for this measurement correspond to an integrated luminosity of 8\pbinv and were recorded with the CMS detector in the year 2010, when the LHC operated at $\sqrt{s}=7$\TeV with low probability of overlapping $\Pp\Pp$ interactions (pileup).

\section{The CMS detector and event reconstruction}

The central feature of the CMS apparatus is a superconducting solenoid of 6\unit{m} internal diameter. Within the field volume are the silicon pixel and strip tracker, the crystal electromagnetic calorimeter (ECAL), and the brass and scintillator hadronic calorimeter (HCAL). Muons are measured in gas-ionization detectors embedded in the steel flux-return yoke outside the solenoid.

The silicon tracker measures charged particles within the pseudorapidity range $\abs{\eta} < 2.5$. It consists of 1440 silicon pixel and 15\,148 silicon strip detector modules. For nonisolated particles of $1 < \pt < 10\GeV$ and $\abs{\eta} < 1.4$, the track resolutions are typically 1.5\% in \pt and 25--90 (45--150)\mum in transverse (longitudinal) impact parameter.
The silicon tracker provides the primary vertex position with $\sim$15\mum resolution for jet events of the type considered in this analysis~\cite{TRK-11-001}.

 In the region $\abs{ \eta }< 1.74$, the HCAL cells have widths of 0.087 in both $\eta$ and azimuth ($\varphi$, in radians). In the $\eta$-$\varphi$ plane, and for $\abs{\eta}< 1.48$, the HCAL cells map onto 5${}\times{}$5 ECAL crystal arrays to form calorimeter towers projecting radially outwards from the nominal interaction point. At larger values of $\abs{ \eta }$, the size of the towers increases and the matching ECAL arrays contain fewer crystals. In addition to the barrel and endcap detectors, CMS has extensive forward calorimetry. The forward component of the hadron calorimeter $(2.9<\abs{\eta}<5.2)$ consists of steel absorbers with embedded radiation-hard quartz fibers, providing fast collection of Cherenkov light.

A more detailed description of the CMS detector, together with a definition of the coordinate system used and the relevant kinematic variables, can be found in Ref.~\cite{Chatrchyan:2008zzk}.

The first level of the CMS trigger system~\cite{CMStrigger}, composed of custom hardware processors, uses information from the calorimeters and muon detectors to select the most interesting events in a fixed time interval of less than 3.2\mus. The high-level trigger processor farm further decreases the event rate from around 100\unit{kHz} to around 400\unit{Hz}, before data storage.

Tracks are reconstructed with the standard iterative algorithm of CMS, which is based on a combinatorial track finder
that uses information from the silicon tracker. To reduce the misidentification rate, tracks are required to pass standard CMS quality criteria, usually referred to as 'high-purity' criteria~\cite{TRK-11-001}. These place requirements on the number of hits, the $\chi^2$ of the track fit, and the degree of compatibility with the hypothesis that the track originates from a vertex reconstructed with the pixel detector. The requirements are functions of the track \pt and $\eta$, as well as the number of layers with a hit. A more detailed discussion on the
combinatorial track finder algorithm and the high-purity track definition can be found in Ref.~\cite{TRK-11-001}.

{\tolerance=1200
The jets are reconstructed using the infrared- and collinear-safe anti-\kt algorithm~\cite{Cacciari:2008gp,Cacciari:2011ma}, with a distance parameter $R=0.5$, starting from the particles identified with the particle-flow method~\cite{PFnew}.
The key feature of the anti-\kt algorithm is the resilience of the jet boundary with respect to soft radiation. This leads to cone-shaped hard jets. Soft jets tend to have more complicated shapes.  The jet momentum is determined as the vector sum of all particle momenta in the jet, and is found in the simulation to be within 5 to 10\% of the true hadron-level momentum over the whole $\pt^{\text{jet}}$ spectrum and detector acceptance. When combining information from the entire detector, the jet energy resolution for jets with $\pt^{\text{jet}} =40$\GeV\,(200\GeV) is about 12\%\,(7\%) for $\abs{\eta^{\text{jet}}}<0.5$ and about 10\% for $4 < \abs{\eta^{\text{jet}}} < 4.5$~\cite{CMS-PAS-JME-10-003}. Jet energy corrections are derived from the simulation, and are confirmed with \textit{in situ} measurements of the energy balance in dijet and photon+jet events~\cite{CMS-PAS-JME-10-010}. No jet energy corrections related to the removal of pileup contributions~\cite{CMS-JME-10-011} are required for the jets studied in this analysis.
\par}

\section{Monte Carlo simulation}
\label{sec:mc}

The simulation of inclusive dijet events is performed using the \PYTHIA6.422 Monte Carlo (MC) event generator~\cite{pythia}. \PYTHIA6 is based on the leading order (LO) DGLAP evolution equations combined with a leading-logarithmic (LL) resummation of soft gluon emission in the parton shower, and uses the Lund string fragmentation model~\cite{Andersson:1983ia} for hadronization. The underlying event in \PYTHIA6 includes particles produced in the fragmentation of minijets from multiple parton interactions (MPI), initial- and final-state radiation, as well as proton remnants. The events were simulated using the Z2* tune~\cite{Chatrchyan:2013gfi}, which was developed to reproduce the CMS underlying event data at center-of-mass energies up to 7\TeV. \PYTHIA6 models the production of diffractive dijets (leading to a final state with a gap-jet-jet topology) and of central diffractive and exclusive dijets (leading to a gap-jet-jet-gap final-state). However, it does not directly generate the jet-gap-jet topology considered here unless a fluctuation in the radiation and hadronization of the parton showers in inclusive dijet production randomly leads to suppressed hadronic activity between the jets.

Jet-gap-jet events are simulated with the default tune of the \HERWIG6.520 generator~\cite{herwig} (switching on CSE production, and switching off all other processes). The \HERWIG6 generator simulates events with hard color-singlet exchange between two partons according to the model by Mueller and Tang~\cite{mt}, which is based on simplified (LL) BFKL calculations. The hadronization process in \HERWIG is based on cluster fragmentation: at the end of the perturbative parton evolution, clusters are built and then decayed into the final-state hadrons. The \HERWIG6 generator does not include any modeling of MPI; they are instead simulated with the \textsc{jimmy} package~\cite{jimmy}. For simplicity, unless stated otherwise, by \HERWIG6 we herafter refer to the combination of this MC generator with \textsc{jimmy}. The \HERWIG6 generator predicts a decrease of the CSE fraction with increasing \pt of the jets, but the Tevatron data show instead the opposite trend~\cite{d01,cdf1}. In the present analysis, the events generated with \HERWIG6 are reweighted with an exponential function, $\exp(\mathrm{b} \, \pt^{\text{jet2}}$) with $\mathrm{b}=0.01\GeV{}^{-1}$, to ensure that the CMS data are reproduced. In the following, this sample of reweighted \HERWIG events will be referred to as the \HERWIG6 sample.

Both \PYTHIA6 and \HERWIG6 use the CTEQ6L1 pa\-ram\-etri\-za\-tion of the proton parton distribution functions \cite{cteq6l}. The simulated events are processed and reconstructed in the same manner as the collision data. A detailed MC simulation of the CMS detector response is performed with the \GEANTfour toolkit \cite{geant}.

\section{Data samples and dijet event selection}
\label{offline}

Three non-overlapping samples of dijet events are used, corresponding to the following three $\pt^{\text{jet}}$ ranges, defined in terms of the \pt of the second leading jet in the dijet system, $\pt^{\text{jet2}}$: 40--60, 60--100, and 100--200\GeV. The first two samples were selected online with dijet triggers with 15 and 30\GeV thresholds on the uncorrected jet $\pt$, respectively, while the third sample was collected with a single jet trigger with uncorrected jet \pt threshold of 70\GeV. This selection maximizes the amount of dijet events for the analysis and ensures high dijet reconstruction efficiency. The triggers for the first two samples were heavily prescaled. The three samples correspond to integrated luminosities of 48, 410, and 8320\nbinv, respectively. The mean number of inelastic $\Pp\Pp$ interactions per bunch crossing (pileup) in each of the three samples is  1.16, 1.17, and 1.60, respectively.

The following conditions are imposed offline on all samples:
\begin{itemize}
\item events are required to contain at least two jets that pass the standard CMS quality criteria \cite{jq};
\item the number of primary vertices with more than zero degrees of freedom in the event, as defined in \cite{TRK-11-001}, is required to be 0 or 1;
\item a primary vertex, if present, is required to be within a longitudinal distance $\abs{z} < 24$ cm from the nominal interaction point;
\item events with long horizontal sections of the pixel tracker traversed by charged particles parallel to the beam (beam-scraping events) are rejected using a dedicated algorithm \cite{bscrap}.
\end{itemize}

In order to allow for a sufficiently wide rapidity gap between the jets, the following conditions are further imposed on the jets:
\begin{itemize}
\item the two leading jets are required to be in the range $1.5<\abs{\eta^\text{jet}}< 4.7$;
\item the two leading jets are required to be in opposite hemispheres: $\eta^\mathrm{jet1} \, \eta^\text{jet2} < 0$.
\end{itemize}

\begin{figure}
\centering
\includegraphics[width=0.48\textwidth]{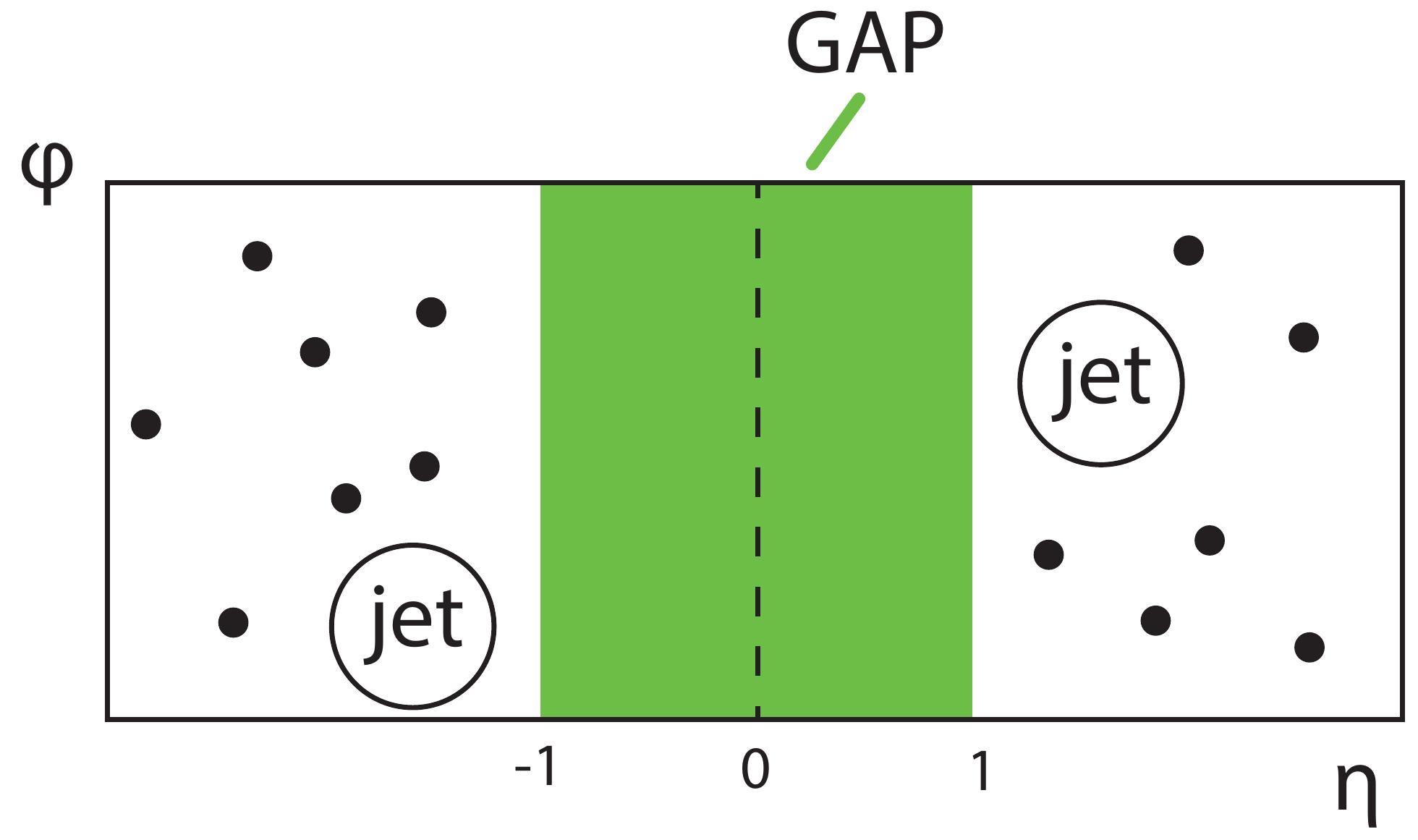}
\caption{Schematic picture of a jet-gap-jet event in the $\varphi$ vs. $\eta$ plane. The circles indicate the two jets reconstructed on each side of the detector, while the dots represent the remaining hadronic activity in the event. The shaded area corresponds to the region of the potential rapidity gap, in which the charged-particle multiplicity is measured (the so-called gap region).}
\label{lego}
\end{figure}

The single- or zero-vertex requirement rejects most of the events with pileup interactions, which can hide an existing rapidity gap. At the same time, it may reject dijet events in which one true primary vertex is wrongly reconstructed as two or more; however, the probability of such badly reconstructed vertices has been checked with the \PYTHIA6~Z2* and \HERWIG6 simulations and found to be negligible. Selecting events with no reconstructed vertices increases the acceptance for signal events in which the two jets are produced outside the tracker coverage. Such events are estimated from the data to contribute about 10\% of all CSE events. According to the simulations the residual fraction of pileup events in the sample is negligible.

There are 6196, 8197, and 9591 events that satisfy the above selection criteria in the $\pt^\text{jet2}= 40$--60, 60--100, and 100--200\GeV jet samples, respectively.

\section{Jet-gap-jet events}
\label{sec:jgj}

The charged-particle multiplicity ($N_\text{tracks}$) in the gap region between the two leading jets (the shaded area in Fig.~\ref{lego}) is used to discriminate between CSE and non-CSE events. The $N_\text{tracks}$ variable is defined as the number of reconstructed particles with $\pt>0.2$\GeV in the interval $\abs{\eta} < 1$. Tracks are required to have a measured \pt with relative uncertainty smaller than 10\% ($\sigma_{\pt}/\pt < 10\%$), which reduces the contribution of tracks from secondary interactions.
 The chosen $\eta$ range ensures a high track reconstruction efficiency and, at the same time, is wide enough to suppress most of the background events with smaller gaps produced via non-CSE fluctuations.

 The separation between the jet axes corresponds to at least three units of $\eta$ (for jets with $\abs{\eta^\text{jet}}> 1.5$ and $\eta^\mathrm{jet1} \, \eta^\text{jet2} < 0$), the minimum gap width typically used in studies of diffractive interactions. For the majority of the events the gap region is far from the edges of jets, which reduces the contamination of soft radiation from the jet shower evolution.

\begin{figure*}
\centering
\includegraphics[width=0.48\textwidth]{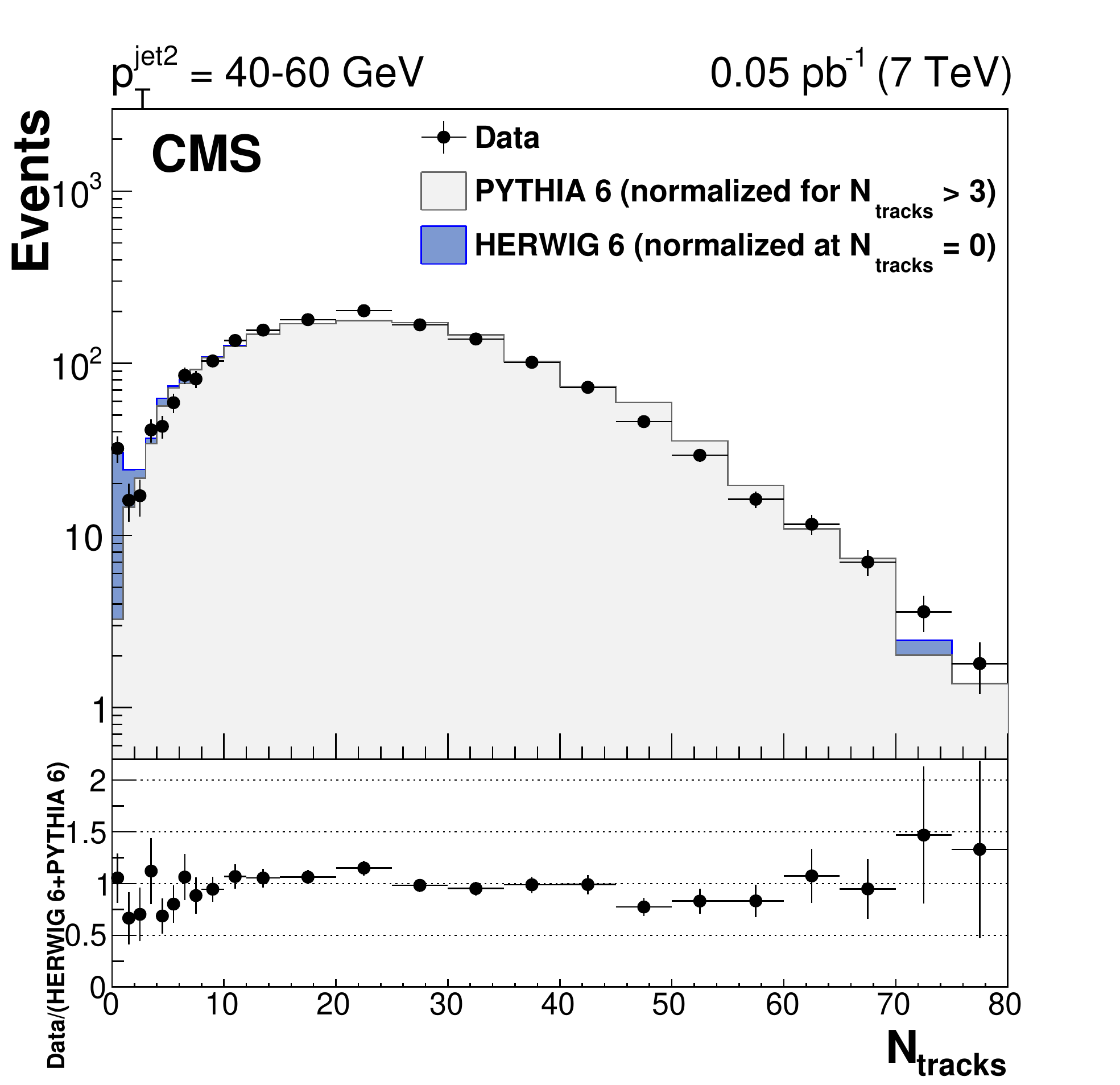}
\includegraphics[width=0.48\textwidth]{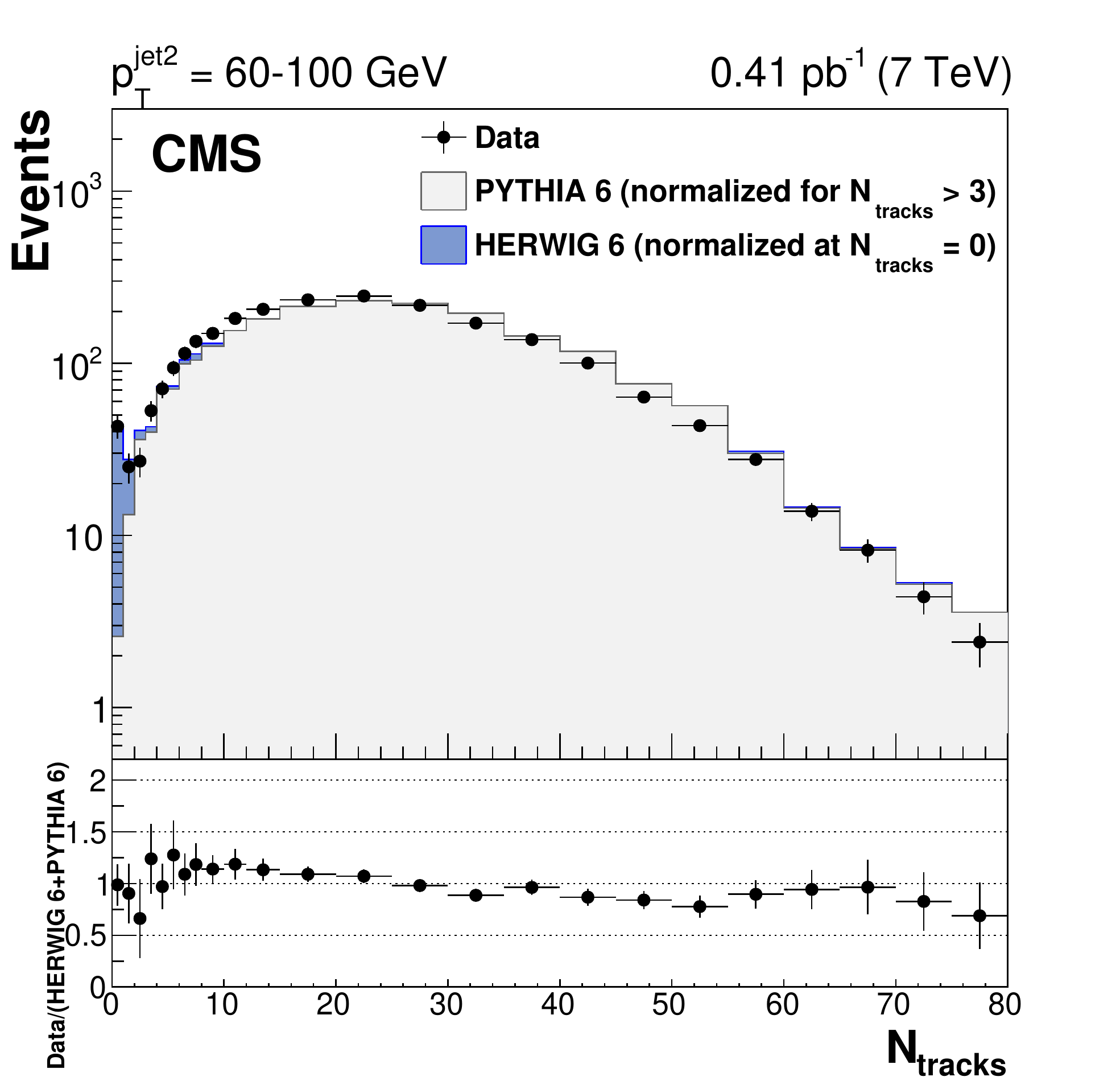}
\includegraphics[width=0.48\textwidth]{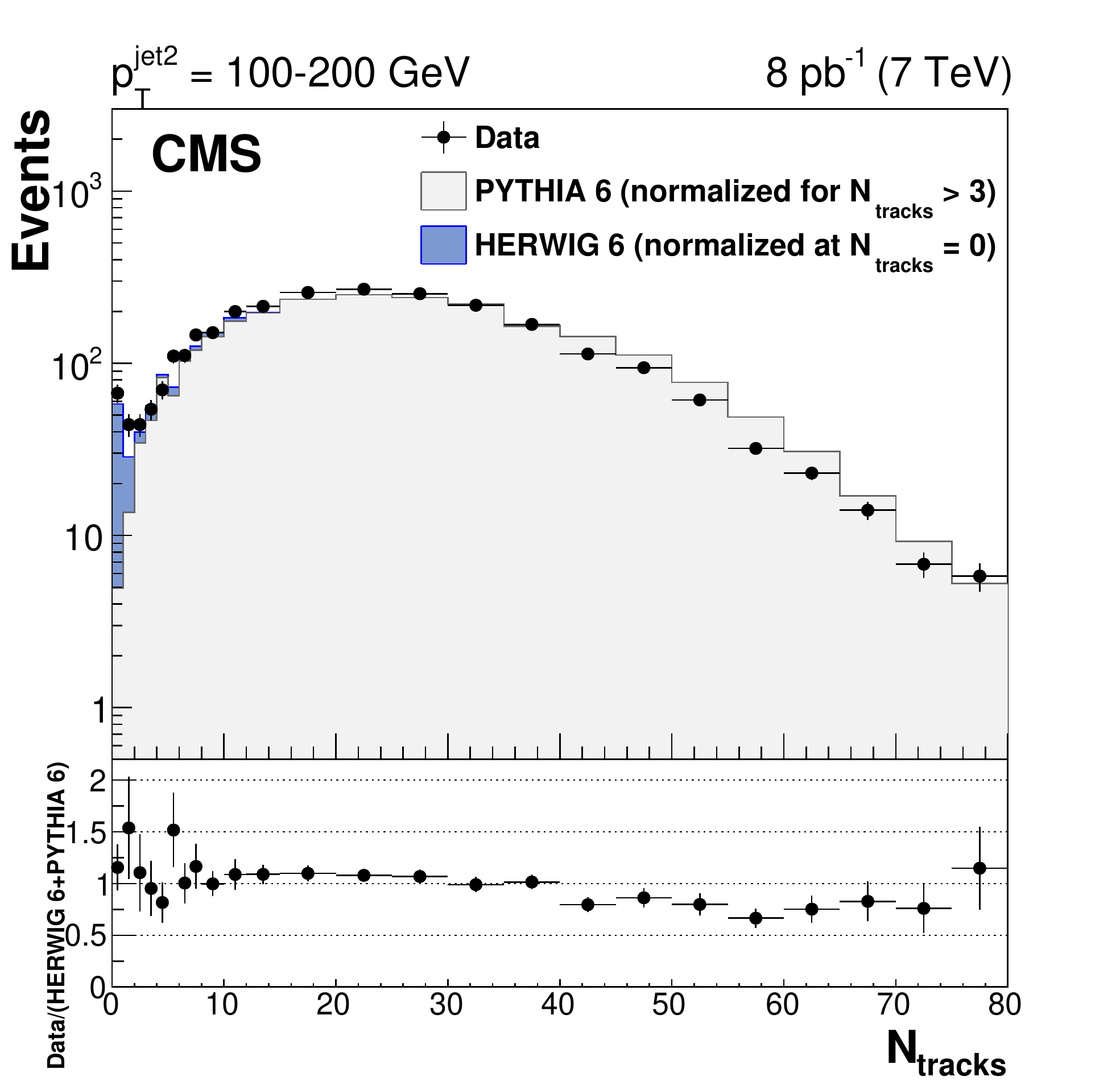}
\caption{Distribution, uncorrected for detector effects, of the number of central tracks between the two leading jets in events with $\pt^\text{jet2}$ = 40--60 (top left), 60--100 (top right), and 100--200 (bottom) \GeV, compared to the predictions of \PYTHIA6 (inclusive dijets) and \HERWIG6 (CSE jet-gap-jet events). The \PYTHIA6 and \HERWIG6 samples are normalized to the number of events measured for $N_\text{tracks}>3$ and  $N_\text{tracks}=0$, respectively. Beneath each plot the ratio of the data yield to the sum of the normalized \HERWIG6 and \PYTHIA6 predictions is shown. The vertical error bars indicate the statistical uncertainties.}
\label{ptmtpl}
\end{figure*}

Figure~\ref{ptmtpl} shows the measured $N_\text{tracks}$ distribution in different $p^\text{jet2}_\mathrm{T}$ bins. In each $\pt^\text{jet2}$ bin, the \PYTHIA6 distribution is normalized to the integral of the number of events measured for $N_\text{tracks}>3$, and the \HERWIG6 predictions are normalized to the number of events with $N_\text{tracks}=0$ measured in the data. The data are satisfactorily described by the \PYTHIA6 simulation, with the exception of the lowest multiplicity bins, in which a large excess of events is observed, consistent with a contribution from CSE events. This excess is well described by the reweighted \HERWIG6 generator, as seen in the data/MC ratio plots.

\begin{figure*}[htbp]
\centering
\includegraphics[width=0.48\textwidth]{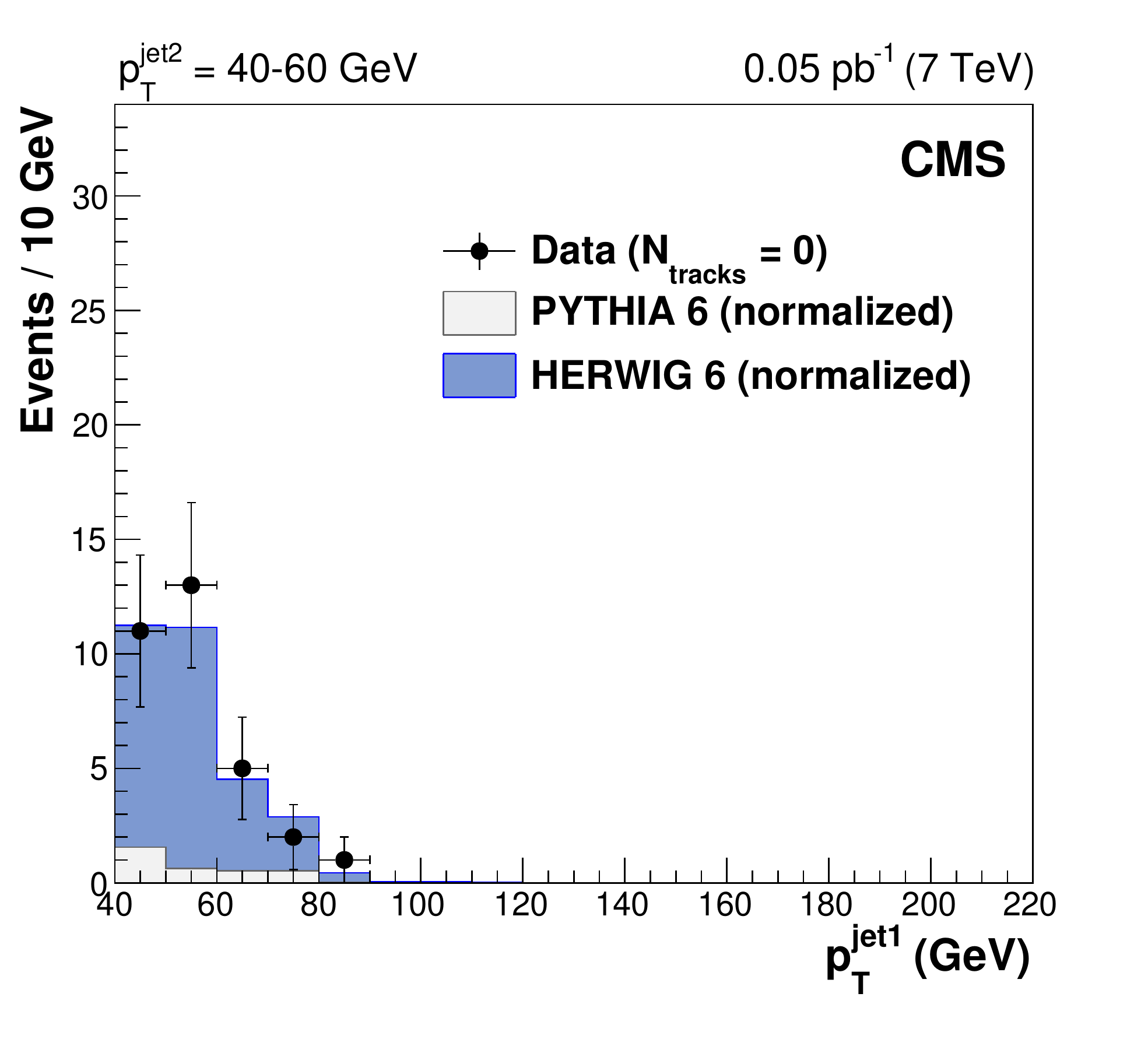}
\includegraphics[width=0.48\textwidth]{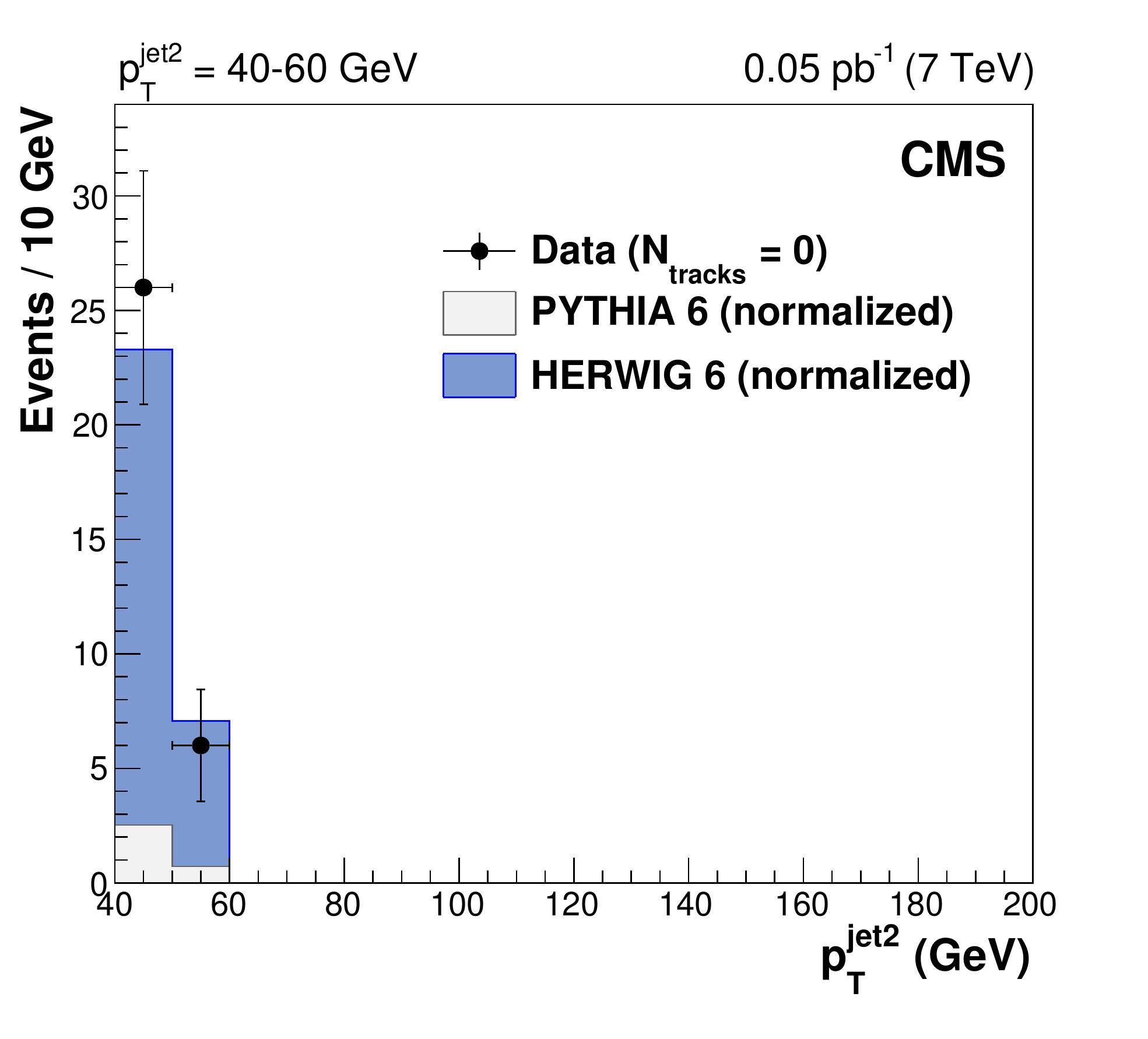}
\includegraphics[width=0.48\textwidth]{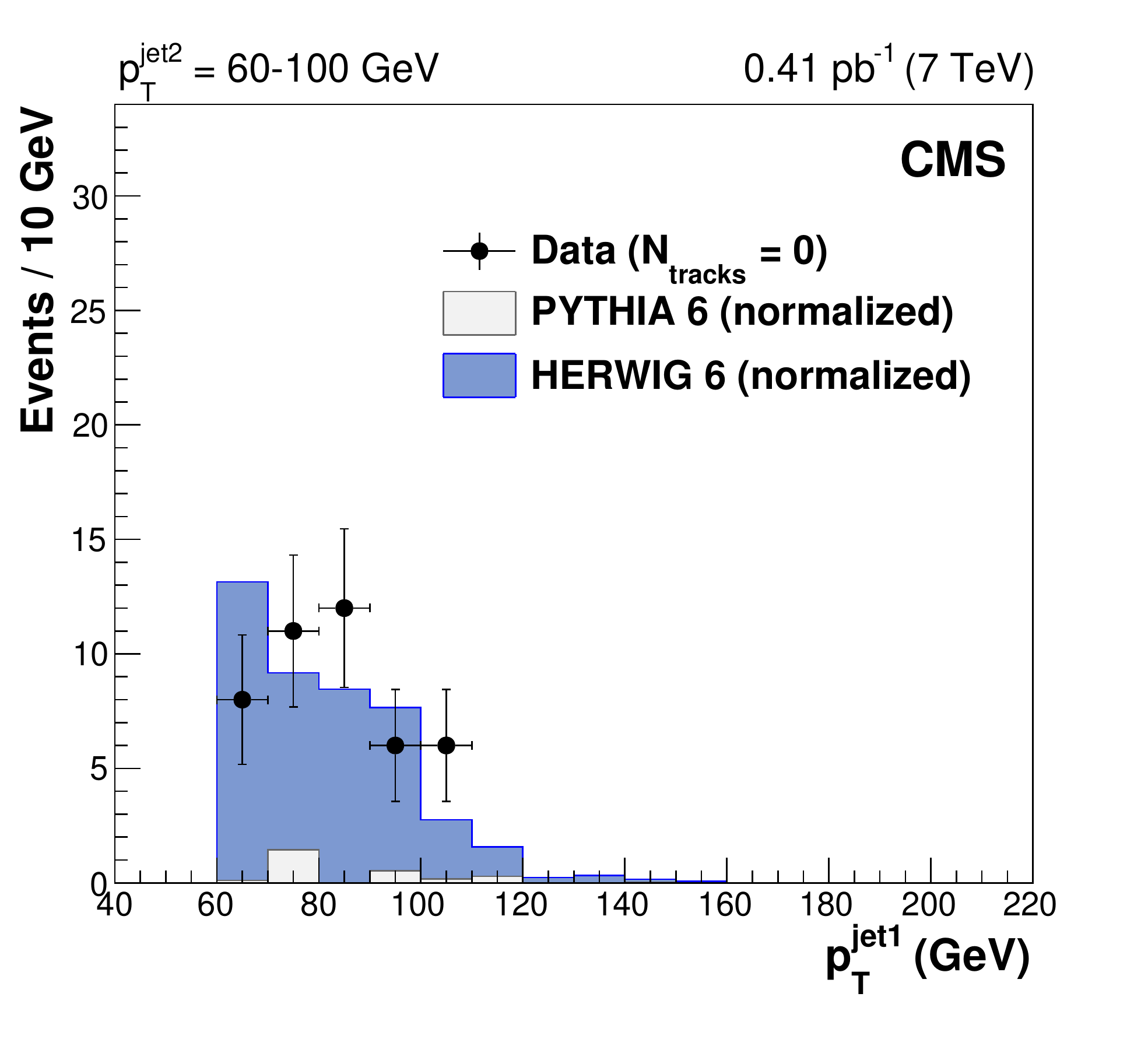}
\includegraphics[width=0.48\textwidth]{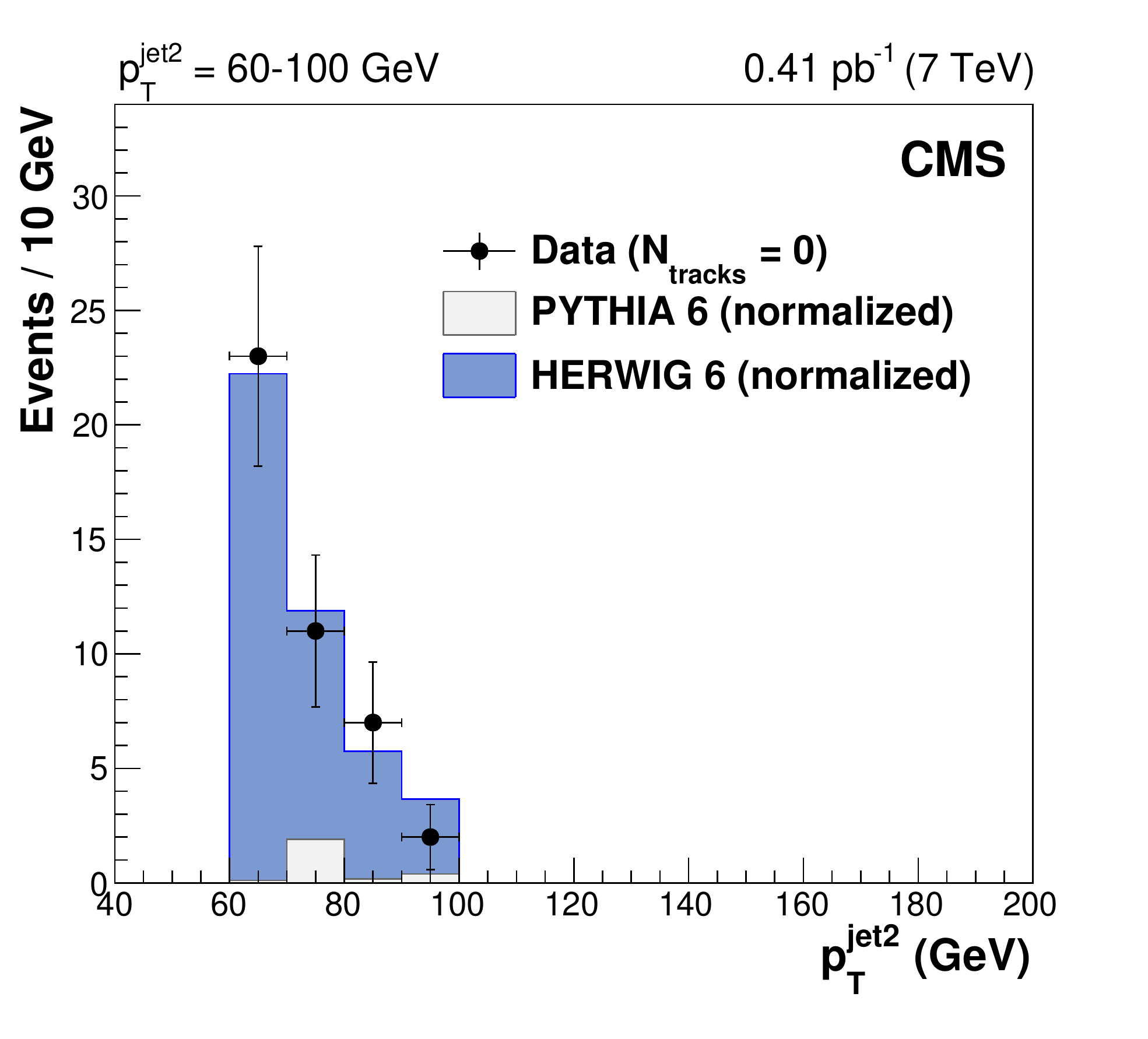}
\includegraphics[width=0.48\textwidth]{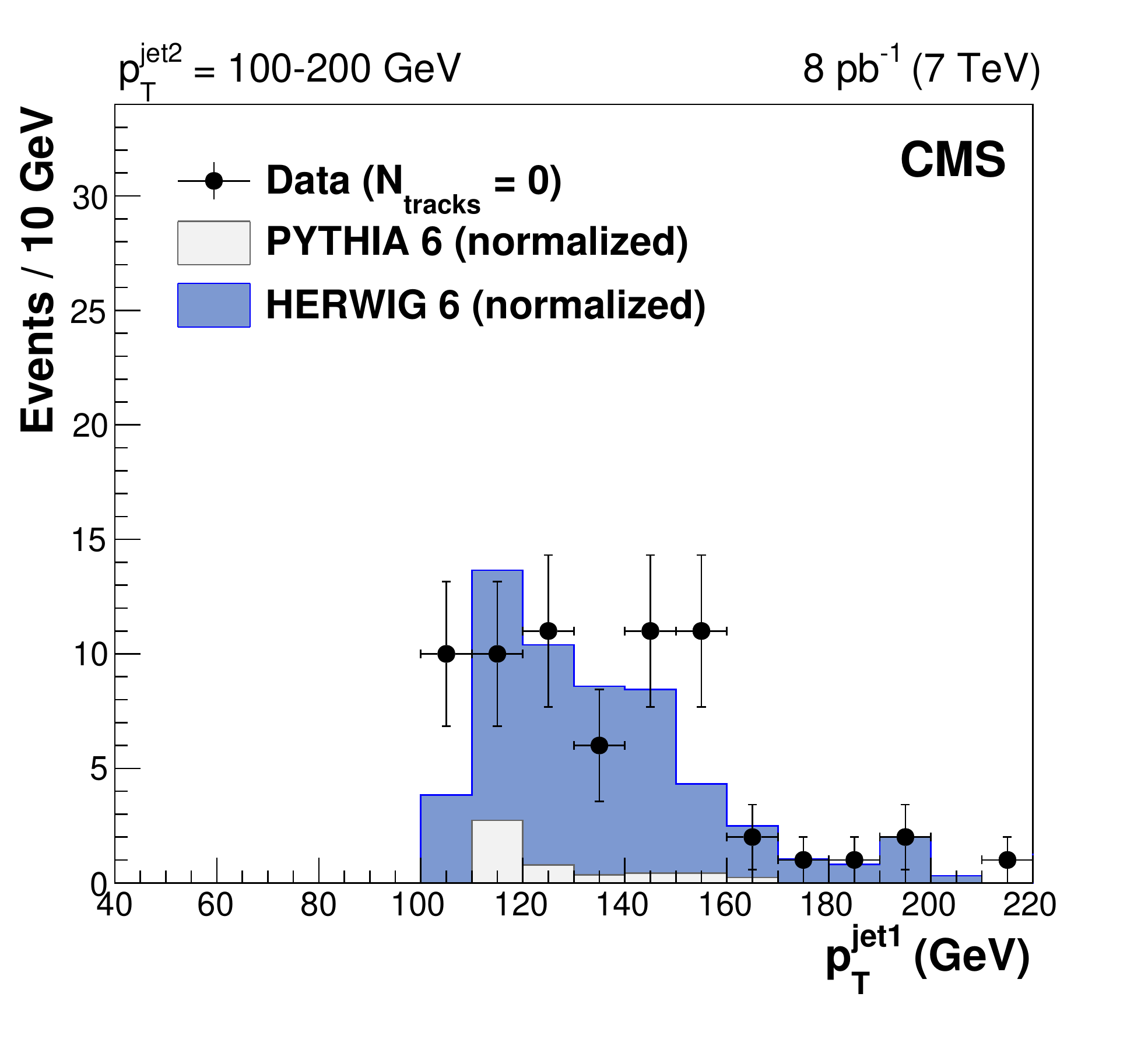}
\includegraphics[width=0.48\textwidth]{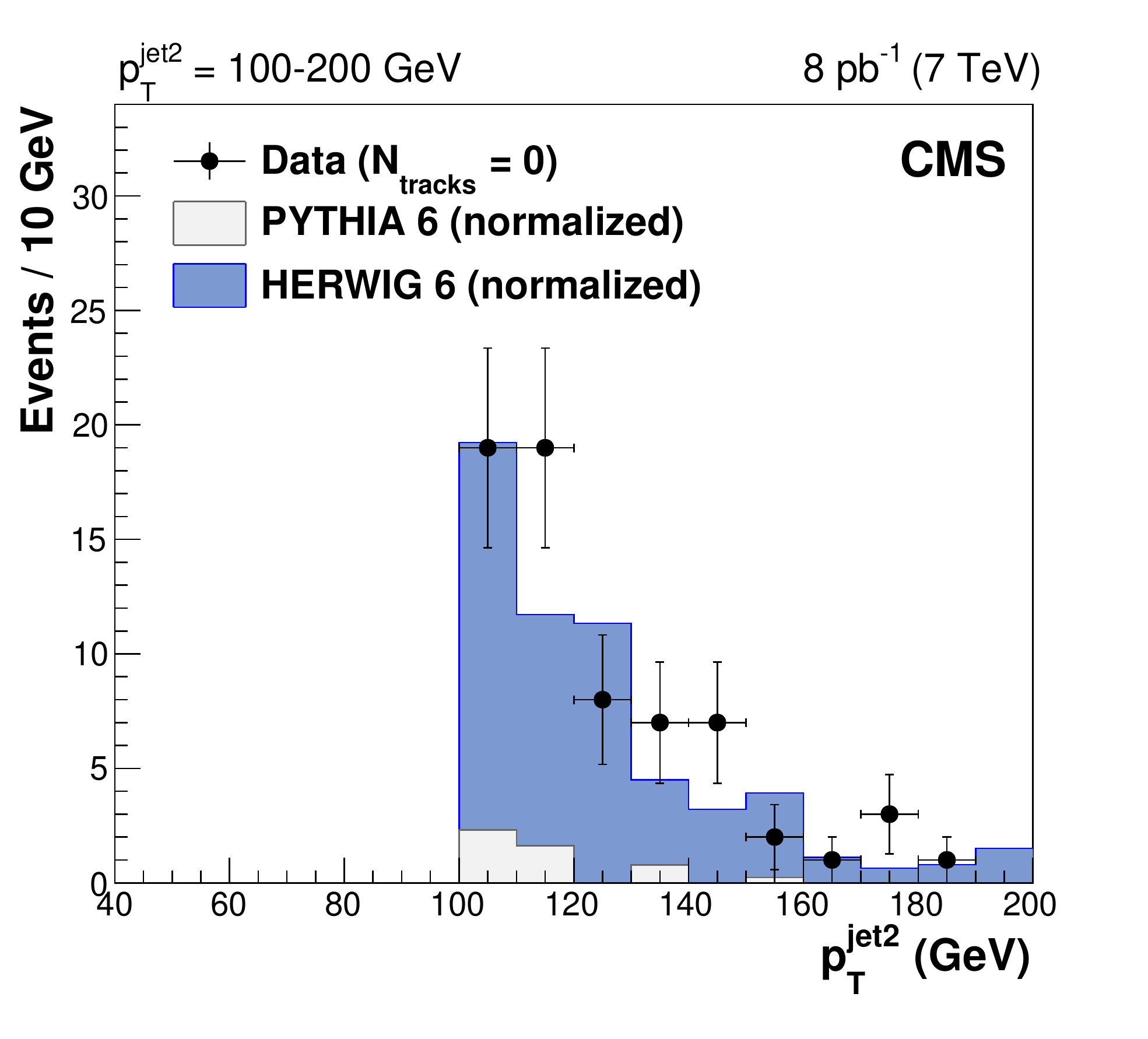}
\caption{Transverse momentum distributions, uncorrected for detector effects, of the leading jet (left) and the second-leading jet (right) in three dijet samples with $\pt^\text{jet2} = 40$--60, 60--100, and 100--200\GeV (from top to bottom) after all selections, for events with no tracks reconstructed in the gap region $\abs{\eta}<1$, compared to predictions of \PYTHIA6 (inclusive dijets) and \HERWIG6 (CSE jet-gap-jet events), normalized as in Fig.~\ref{ptmtpl}. The error bars indicate the statistical uncertainties.}
\label{ptj12}
\end{figure*}

\begin{figure*}[t!]
\centering
\includegraphics[width=0.99\textwidth]{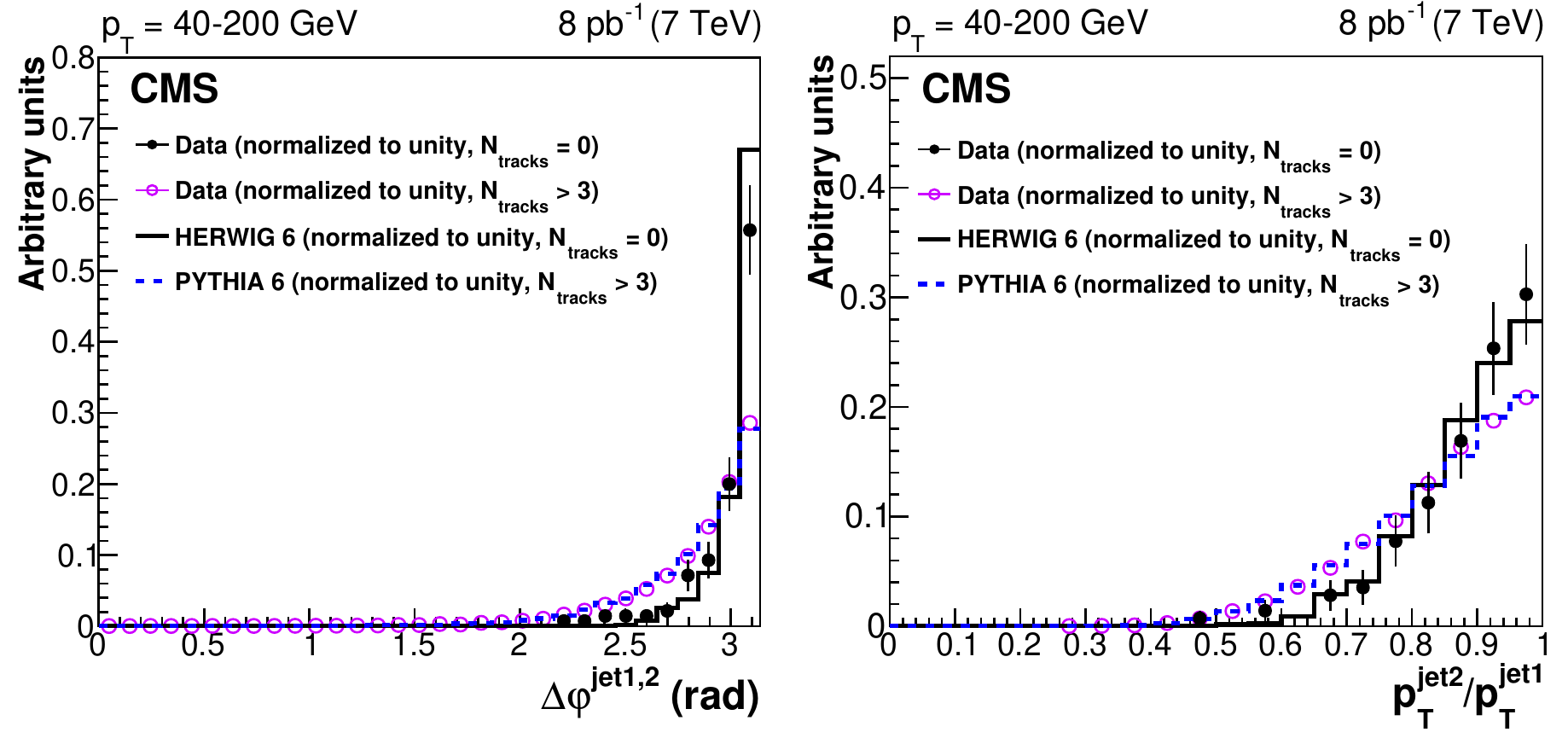}
\caption{Distributions, uncorrected for detector effects, of the azimuthal angle $\Delta \varphi^\mathrm{jet1,2}$ between the two leading jets (left) and the ratio $\pt^\text{jet2}/\pt^\mathrm{jet1}$  of the second-leading jet \pt to the leading jet \pt (right) for events after all selections, with no tracks ($N_\text{tracks} = 0$, full circles) or more than three tracks ($N_\text{tracks}> 3$,  open circles) reconstructed in the $\abs{\eta}<1$ region, compared with the MC predictions. The distributions are summed over the three $p^\text{jet2}_\mathrm{T}$ bins used in the analysis and normalized to unity for shape comparison.}
\label{dphir}
\end{figure*}

The leading and the second-leading jet \pt spectra for events with no tracks reconstructed in the gap region $\abs{\eta}<1$ are presented in Fig.~\ref{ptj12}. The data, plotted in bins of $p^\text{jet2}_\mathrm{T}$, are reproduced by the normalized \HERWIG6 CSE events. A very small contribution from \PYTHIA6 events can be explained by fluctuations in the hadronization of (non-CSE) inclusive dijet events, with no particles or only neutral particles produced inside the gap region. Figure~\ref{dphir} shows the distributions of the azimuthal angle $\Delta \varphi^\mathrm{jet1,2}$ between the jets (left), and of the ratio of the second-leading jet \pt to the leading jet $\pt$, $\pt^\text{jet2}/\pt^\mathrm{jet1}$ (right). The data, shown separately for events with no tracks and with more than three tracks reconstructed in the $\abs{\eta}<1$ region, are well described by the normalized simulations, which are dominated by CSE (\HERWIG6) and non-CSE (\PYTHIA6) events, respectively. The peaks in the distributions at $\Delta \varphi^\mathrm{jet1,2}=\pi$ and  $\pt^\text{jet2}/\pt^\mathrm{jet1}=1$ are narrower for events with no tracks, reflecting the fact that the CSE dijets are more balanced in azimuthal angle and momentum than the non-CSE ones, because of the extra radiation in the latter.

In order to quantify the contribution from CSE events, we measure the CSE fraction, $f_\mathrm{CSE}$, defined as
\begin{equation}
f_\mathrm{CSE}=\frac{N_\text{events}^\mathrm{F}-N_\text{non-CSE}^\mathrm{F}}{N_\text{events}},
\label{eq-1}
\end{equation}
where $N_\text{events}^\mathrm{F}$ is the number of events in the first bins of the multiplicity distribution ($N_\text{tracks}<2$ or 3, as explained later in this Section), $N_\text{non-CSE}^\mathrm{F}$ is the estimated number of events in these bins originating from non-CSE events, and $N_\text{events}$ is the total number of events considered. The $f_\mathrm{CSE}$ fraction defined in this way is not sensitive to the trigger efficiencies and jet reconstruction uncertainties as they cancel in the ratio. While the extraction of $N_\text{events}^\mathrm{F}$ and $N_\text{events}$ is straightforward (event counting), the estimation of $N_\text{non-CSE}^\mathrm{F}$ requires modeling of the non-CSE contributions, for which two data-driven approaches are considered.

\begin{figure*}[htbp]
\centering
\includegraphics[width=0.48\textwidth]{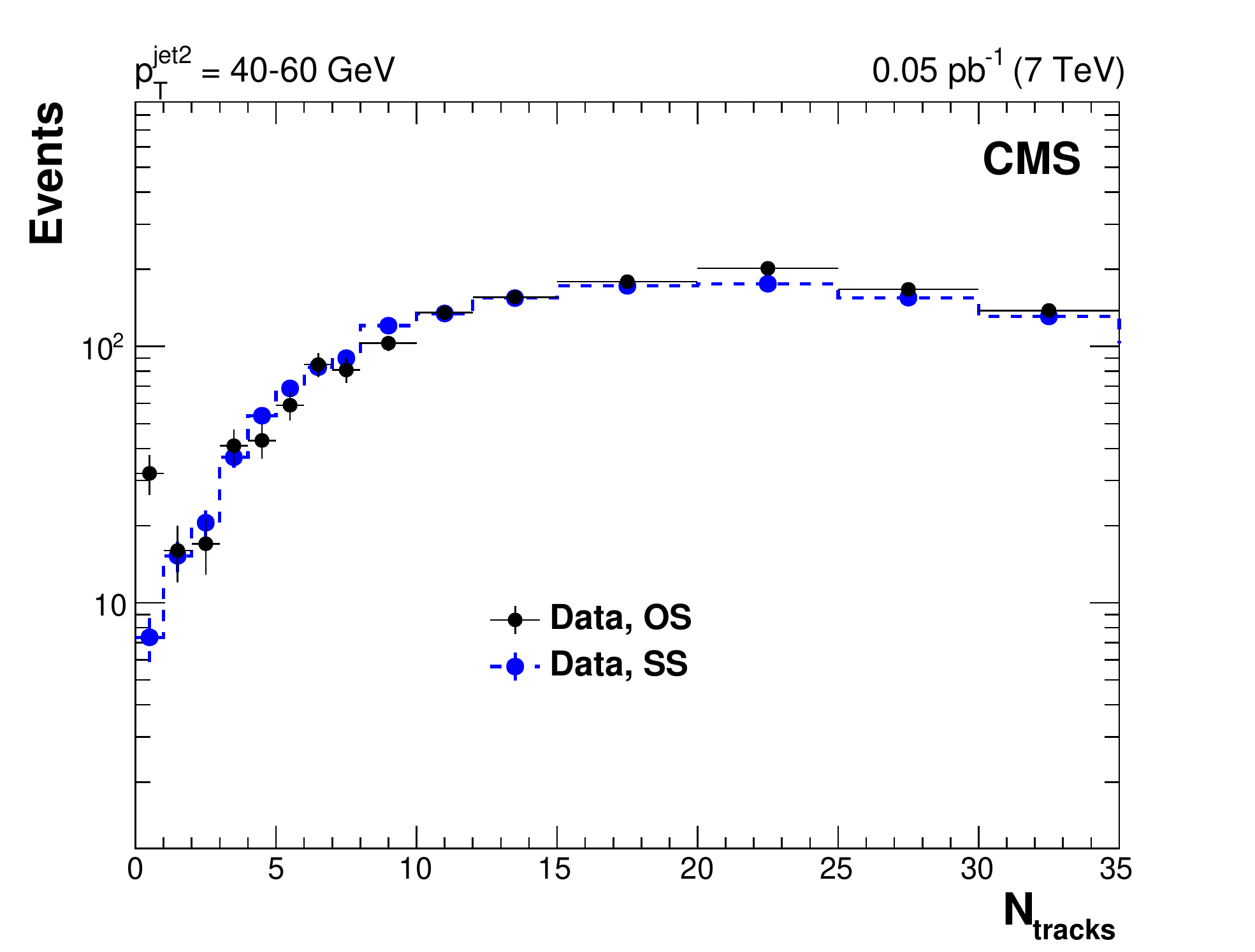}
\includegraphics[width=0.48\textwidth]{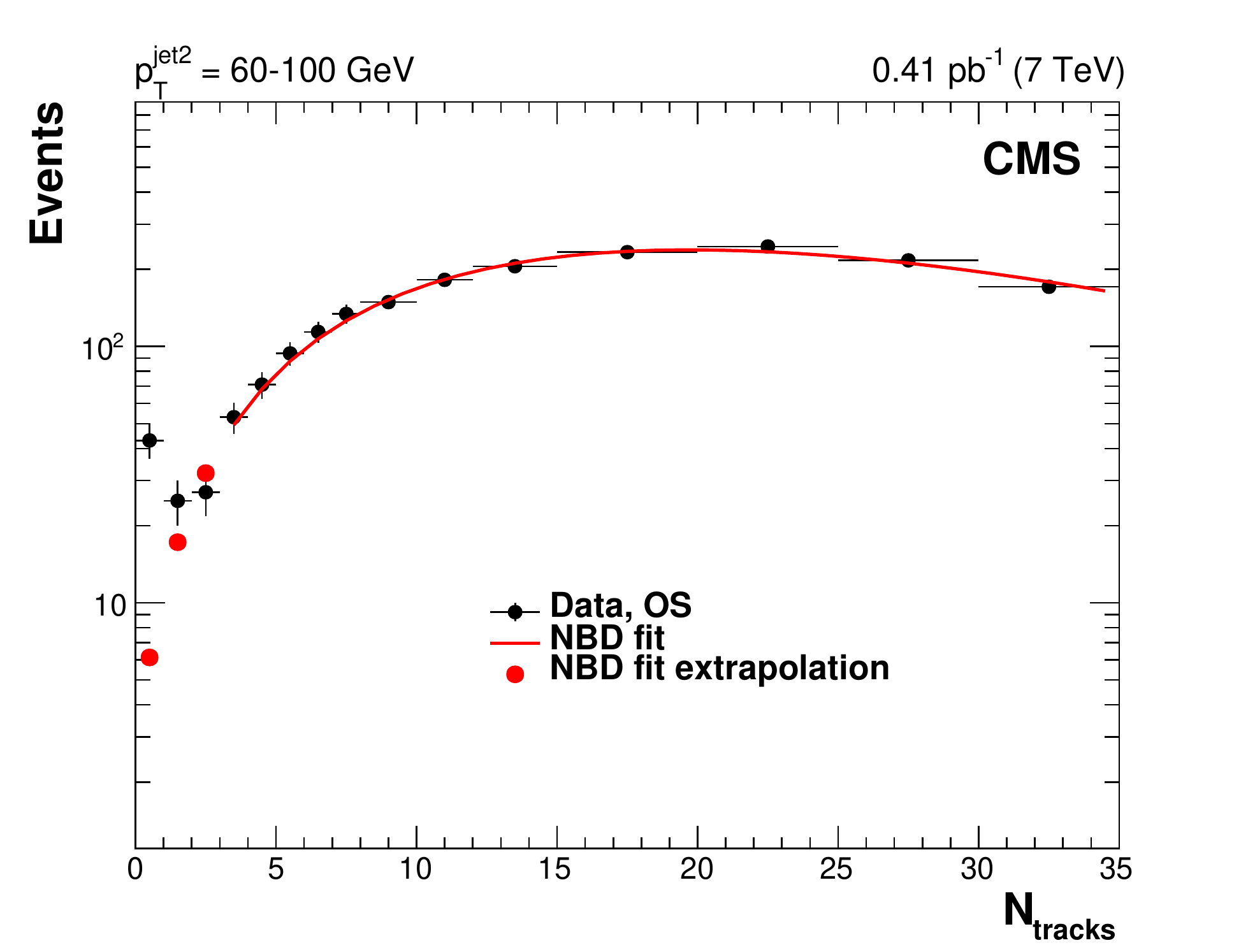}
\includegraphics[width=0.48\textwidth]{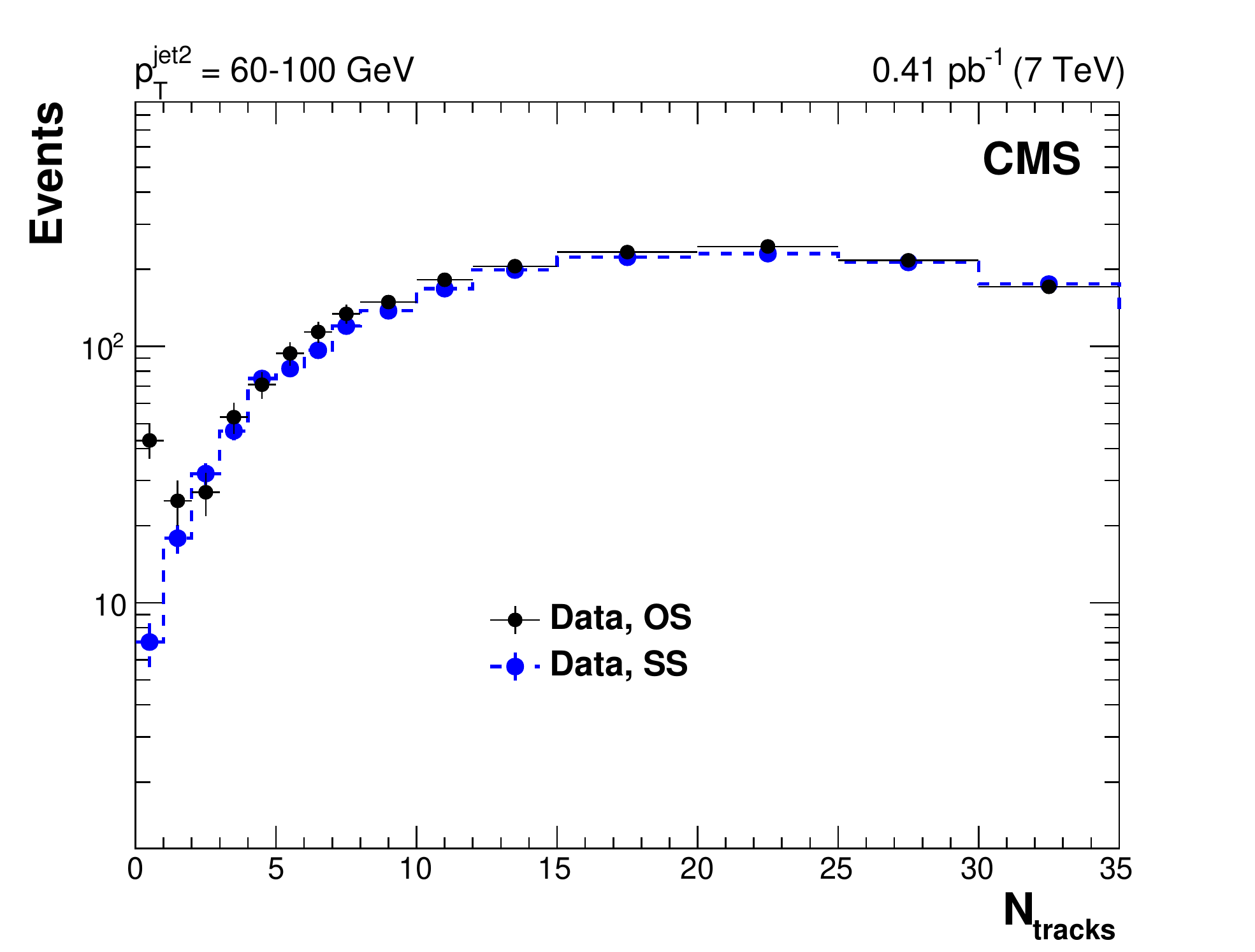}
\includegraphics[width=0.48\textwidth]{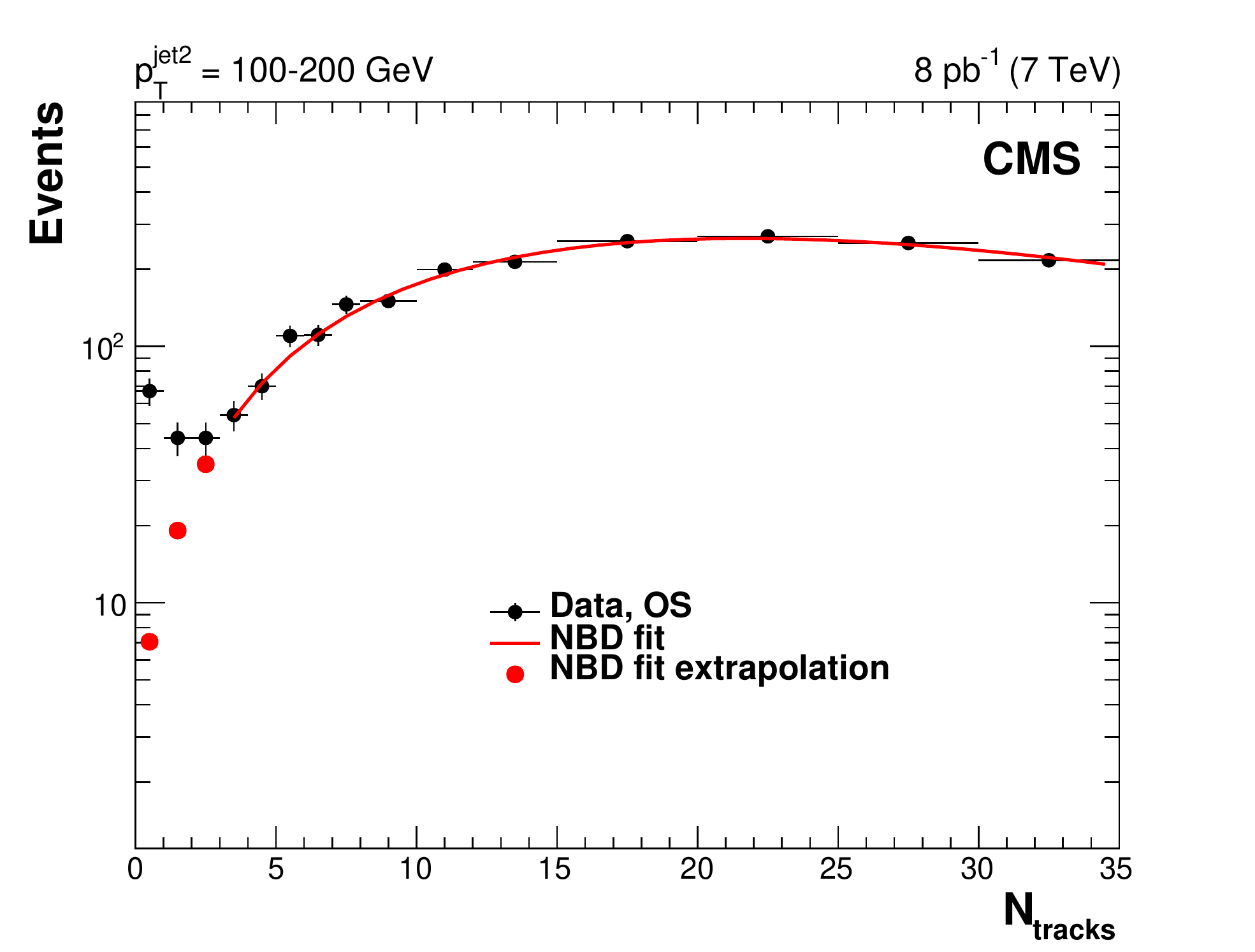}
\includegraphics[width=0.48\textwidth]{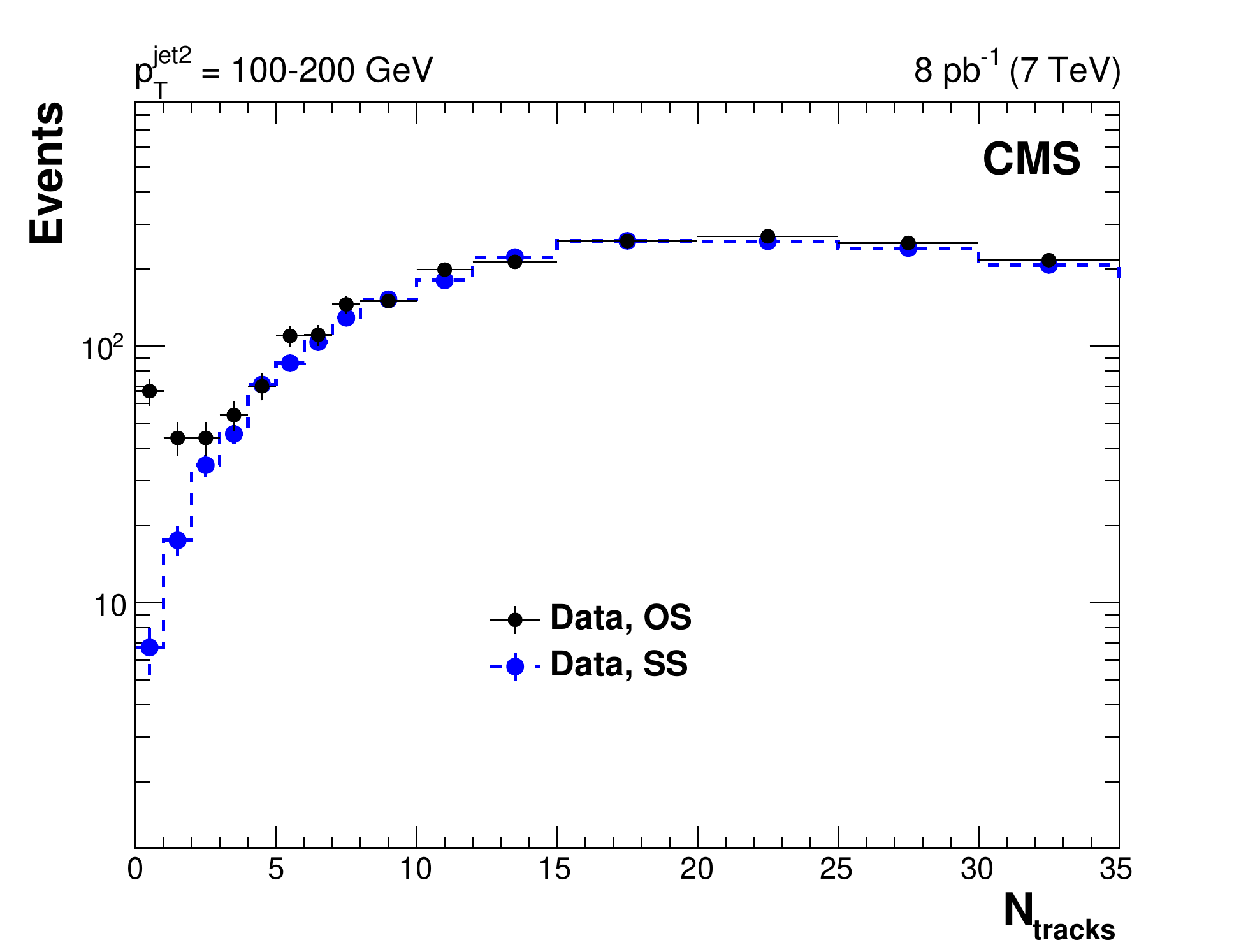}
\includegraphics[width=0.48\textwidth]{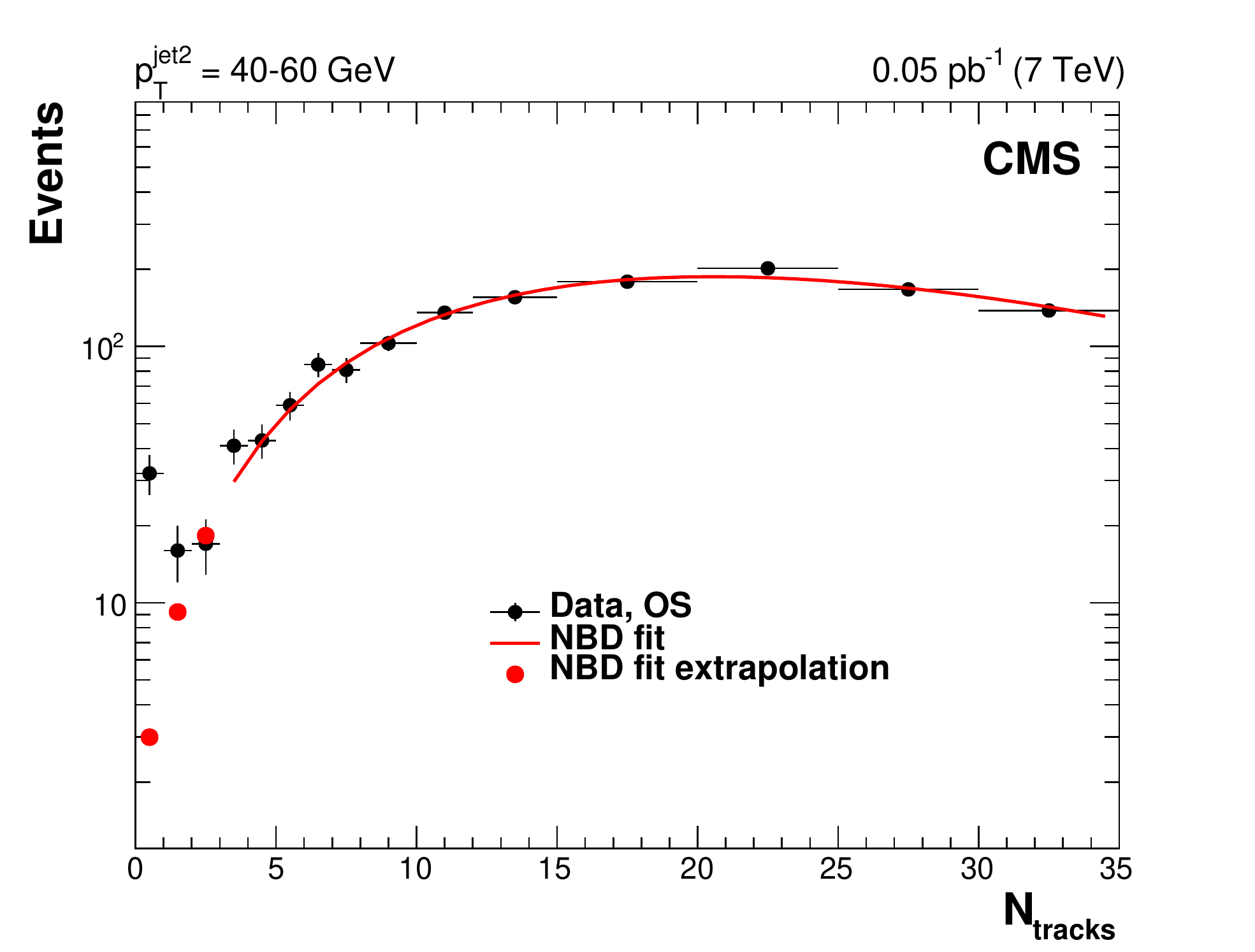}
\caption{Distribution, uncorrected for detector effects, of the number of central tracks in opposite-side (OS) dijet events (black circles) with $\pt^\text{jet2}$ = 40--60 (top), 60--100 (middle), and 100--200\GeV (bottom), plotted (left) together with the $N_\text{tracks}$ distribution of same-side (SS) dijet events (blue circles), and fitted to a NBD function (right).}
\label{ssnbd}
\end{figure*}

In the first approach, the shape of the $N_\text{tracks}$ distribution for background events is obtained from a sample in which the two leading jets are produced on the same side of the CMS detector (same side, or SS, sample, with jets satisfying the selection $\abs{\eta^\text{jet}}> 1.5$  and $\eta^\mathrm{jet1} \, \eta^\text{jet2} > 0$).  For the nominal sample defined in Section~\ref{offline} (opposite side, or OS, sample, with two jets produced on opposite sides of the CMS detector), the gap region $\abs{\eta} < 1$ mainly contains particles originating from the hard scattering, while for the SS sample it is dominated by particles originating from the underlying event. This difference is reflected in the $N_\text{tracks}$ distributions: whereas the shapes of the distributions are similar for the SS and OS samples, the mean $N_\text{tracks}$ value in the SS sample is slightly lower. In order to minimize the difference between the average $N_\text{tracks}$ values of the two samples, the gap region for the SS sample is enlarged to  $\abs{\eta} < 1.2$, in agreement with the range reported by the CDF Collaboration~\cite{cdf3}. The adjusted multiplicity distribution in the SS sample (Fig.~\ref{ssnbd} left) is normalized to the one in the OS sample for $N_\text{tracks} > 3$, and the number of events in the first bins is taken as an estimate of the background.

The second method is based on the fit of the $N_\text{tracks}$ distribution with a negative binomial distribution (NBD), which was first used to describe charged-particle multiplicity distributions by the UA5 Collaboration~\cite{ua5} at energies up to $\sqrt{s} = 546$\GeV. Later, it was observed that the NBD fit reproduces less well the tails of the particle multiplicity at higher center-of-mass energies (deviations were reported at $\sqrt{s} = 900$\GeV by UA5, and later at Tevatron and LHC energies~\cite{ua5failstart, d02, alice}). This issue is largely avoided when one restricts the NBD fit to the region around the mean of the distribution. The fit used in this analysis starts at $N_\text{tracks} = 3$, where the CSE signal to background ratio is expected to be negligible, and ends at $N_\text{tracks} = 35$, slightly above the maximum of the distribution. The extrapolation of the fit to the first multiplicity bins provides an estimate of the non-CSE background. The results of the NBD fits are shown in Fig.~\ref{ssnbd} (right). To check the performance of the method, the fit is repeated on the SS sample in the range $3 \le N_\text{tracks} \le 35$. The extrapolation of the fit to the $N_\text{tracks} < 3$ region agrees with the number of events observed in the SS sample data, which confirms the validity of this approach.

The numbers of background events obtained with the two methods described above agree within statistical uncertainties, with the results of the NBD fit being slightly lower. Since the SS method cannot be used to estimate the background in bins of $\Delta\eta_\mathrm{jj}$ between the jets (because of the smaller $\Delta\eta_\mathrm{jj}$ values than in the OS sample), the NBD fit is chosen as the main background determination method in this analysis. The method involving the SS sample is used as a systematic check, as discussed in the next section.
The non-CSE background contributes about 10--15\% of the events in the $0^\mathrm{th}$ bin of the multiplicity distribution, about 25--35\% in the first two multiplicity bins, and about 40--60\% when the signal is integrated over the first three multiplicity bins.

\begin{figure}
\centering
\includegraphics[width=0.48\textwidth]{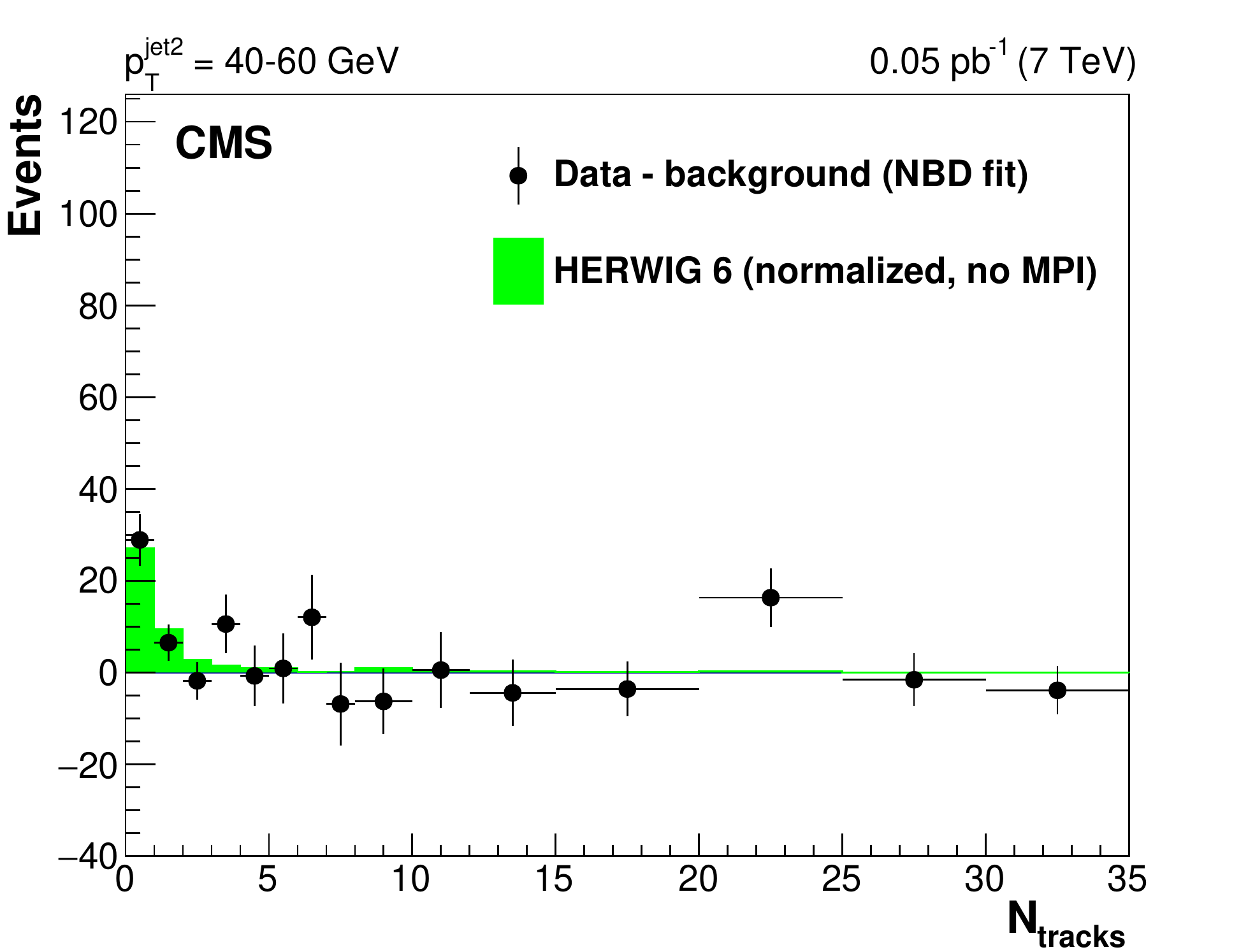}
\includegraphics[width=0.48\textwidth]{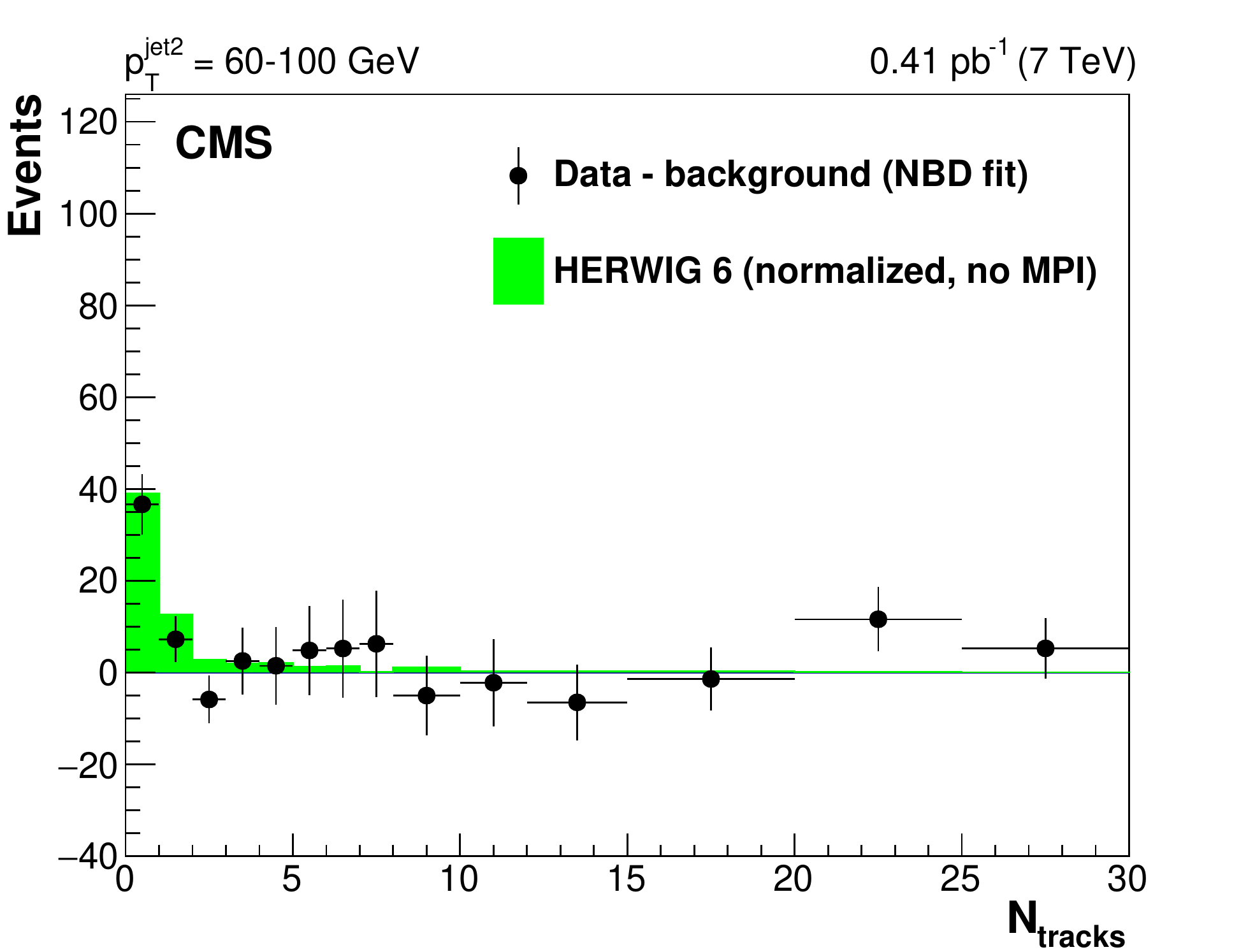}
\includegraphics[width=0.48\textwidth]{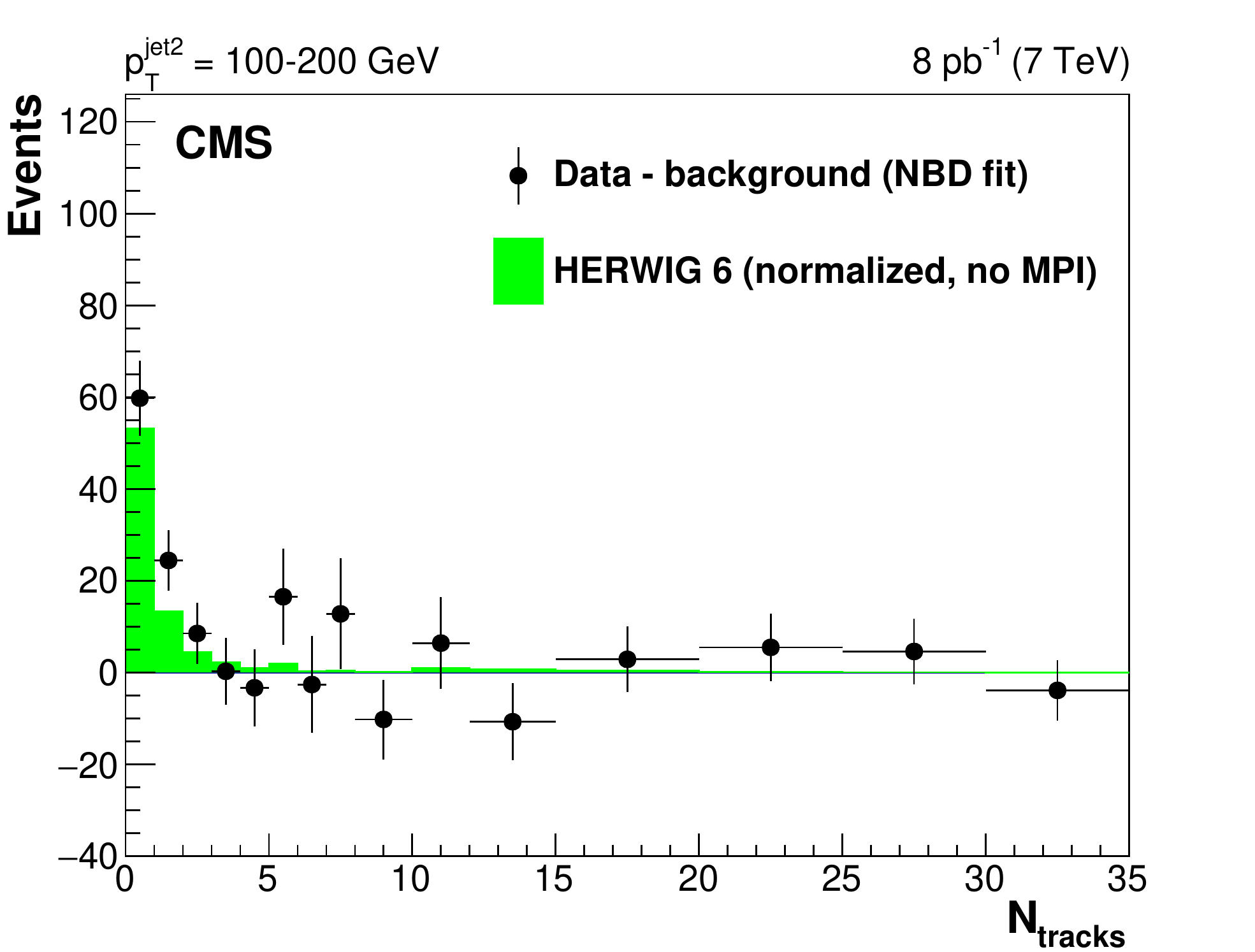}
\caption{Background-subtracted central track multiplicity distributions, uncorrected for detector effects, in the three bins of $\pt^\text{jet2}$, compared to the \HERWIG6 predictions without underlying event simulation (``no MPI''), normalized as in Fig.~\ref{ptmtpl}. The background is estimated from the NBD fit to the data in the $3 \le N_\text{tracks} \le 35$ range, extrapolated to the lowest multiplicity bins.}
\label{subpt}
\end{figure}

Figure~\ref{subpt} shows the track multiplicity distribution in the three bins of $\pt^\text{jet2}$ after subtracting the non-CSE background. A clear excess in the lowest bins is observed over a flat continuum, in agreement with the normalized predictions from a \HERWIG6 subsample with jet-gap-jet events only (no additional MPI); the jet-gap-jet events with additional MPI producing tracks in the rapidity gap are part of the background subtracted from the track multiplicity distributions, and are not included in the figure. In the region of the excess (CSE signal region), most events are in the $0^\text{th}$ bin, with smaller contributions from events with one or two tracks reconstructed in the gap region. These tracks originate from the jets but are reconstructed outside of the jet cone, and their contribution is larger in the highest $\pt^\text{jet2}$ bin, for which jets tend to have a higher multiplicity and to be produced more centrally (closer to the gap). We use the $N_\text{tracks}<2$ region to extract the CSE signal in the lowest and medium $\pt^\text{jet2}$ bins, and the $N_\text{tracks}<3$ region to extract the CSE signal in the highest $\pt^\text{jet2}$ bin.

The CSE fractions are obtained from the data using Eq.~(\ref{eq-1}), with the different terms in this formula uncorrected for detector effects. No unfolding of the data is necessary since the effects of resolution and migration of the dijet variables cancel in the $f_\mathrm{CSE}$ ratio. In addition, the number of jet--gap--jet events extracted in the numerator of Eq.~(\ref{eq-1}) does not depend on the track reconstruction efficiency; the latter only influences the non-CSE background count, which is subtracted from the data. Studies with simulated events show that the results do not change, within uncertainties, if the hadron-level variables are used. For the latter, stable particles (with lifetime $\tau$ such that $c\tau > 10$\unit{mm}) are used both for the jet reconstruction and for the extraction of the $N_\text{tracks}$ variable.

\section{Systematic uncertainties}
\label{sec-syst}

The systematic uncertainties in the $f_\mathrm{CSE}$ extraction are estimated by modifying the selection criteria and the analysis procedure. The following sources of systematic uncertainty are taken into account:

\begin{itemize}
\item {Jet energy scale (JES):} the \pt of each jet in an event is varied up and down according to the formula $\pt^\text{jet, new}= \pt^{\text{jet}} \pm \mathrm{u}(\pt^{\text{jet}}, \eta^{\text{jet}})$, where $\mathrm{u}(\pt^{\text{jet}}, \eta^{\text{jet}})$ is the JES uncertainty, which increases at lower (higher) values of $\pt^{\text{jet}}$ ($\eta^{\text{jet}}$)~\cite{jq}. After changing the \pt of the jets, they are reordered in $\pt^\text{jet, new}$, and the analysis is repeated using the two highest $\pt^\text{jet, new}$ jets. The average difference of the results obtained for the positive and negative variations relative to the nominal result is taken as an estimate of the uncertainty associated with the JES.
\item {Track quality:} the track multiplicity distributions are redetermined after relaxing the track quality criteria \cite{TRK-11-001}, in order to study the effect of variations in the track finding algorithm.  The symmetrized difference between the results obtained with the relaxed and nominal conditions is taken as an estimate of the uncertainty.
\item {Background subtraction:} the number of background events in the first bins of the $N_\text{tracks}$ distribution is estimated from data, based on the SS sample introduced in Section~\ref{sec:jgj}. The symmetrized difference of the results with respect to those found with the nominal method, based on the NBD fit, is taken as an estimate of the corresponding uncertainty. For the  measurement of $f_\mathrm{CSE}$ as a function of $\Delta\eta_\mathrm{jj}$ in bins of $\pt^\text{jet2}$, the average uncertainty in the $\pt^\text{jet2}$ bin is used in each $\Delta\eta_\mathrm{jj}$ bin.
\end{itemize}

\begin{table*}
\centering
 \topcaption{Percent systematic (individual, and total) and statistical uncertainties of the CSE fraction in the three bins of $\pt^\text{jet2}$.}
\label{syspttab}
\begin{tabular}{lccc}
\hline
Source                 & 40--60\GeV & 60--100\GeV & 100--200\GeV\\
\hline
Jet energy scale       & $\pm$5.1   & $\pm$6.7 & $\pm$2.1  \\
Tracks quality         & $\pm$0.3   & $\pm$1.3 & $\pm$0.4  \\
Background subtraction & $\pm$14.1& $\pm$0.9 & $\pm$1.9\\[\cmsTabSkip]
Total systematic       & $\pm$15.0 & $\pm$6.9 & $\pm$2.8\\
Statistical            & $\pm$23 & $\pm$22 & $\pm$15 \\
\hline
\end{tabular}
\end{table*}

The total systematic uncertainty is calculated as the quadratic sum of the individual contributions. The effect of each systematic source and the total systematic uncertainty are also given in Table~\ref{syspttab}, for each of the $\pt^\text{jet2}$ bins. In this analysis, the systematic uncertainties are smaller than the statistical ones.

As a check of the sensitivity of the results to the definition of the hadronic activity in the gap region, the track multiplicity distributions are redetermined after increasing the lower limit of the track \pt from 0.2\GeV to 0.25\GeV. The results agree within a few percent with the nominal ones, implying no strong dependence on the hadronic activity definition. This observation is in accordance with the results of the D0 experiment~\cite{d03} using calorimeter towers, in which consistent values of the $f_\mathrm{CSE}$ fraction were obtained for tower transverse energy thresholds of 0.15\GeV, 0.2\GeV and 0.25\GeV. Likewise, in the CDF analysis~\cite{cdf2} consistent results were obtained based on track multiplicities ($\pt>0.3$\GeV) and calorimeter tower multiplicities ($\et>0.2$\GeV). In the present analysis, neutral particles are not included in the multiplicity calculation because of the relatively high transverse energy thresholds required above calorimeter noise, about 0.5\GeV for photons and 2\GeV for neutral hadrons, compared to the much lower 0.2\GeV value for charged tracks.

\section{Results}
\label{measpt}

\begin{table}
\centering
 \topcaption{Measured values of $f_\mathrm{CSE}$ as a function of $\pt^\text{jet2}$. The first and second uncertainties correspond to the statistical and systematic components, respectively. The mean values of $\pt^\text{jet2}$ in the bin are also given.}
\label{pttab}
\begin{tabular}{ccc}
 \hline
 $\pt^\text{jet2}$ range (\GeVns{}) & $\langle \pt^\text{jet2} \rangle$ (\GeVns{})& $f_\mathrm{CSE}$ (\%) \\
\hline
\pmrule 40--60              & 46.6  & $0.57\pm 0.13 \pm 0.09$   \\
\pmrule 60--100             & 71.2  & $0.54\pm 0.12 \pm 0.04$   \\
\pmrule 100--200            &120.1  & $0.97\pm 0.15 \pm 0.03$   \\
\hline
\end{tabular}
\end{table}

\begin{figure} \centering
\includegraphics[width=\cmsFigWidth]{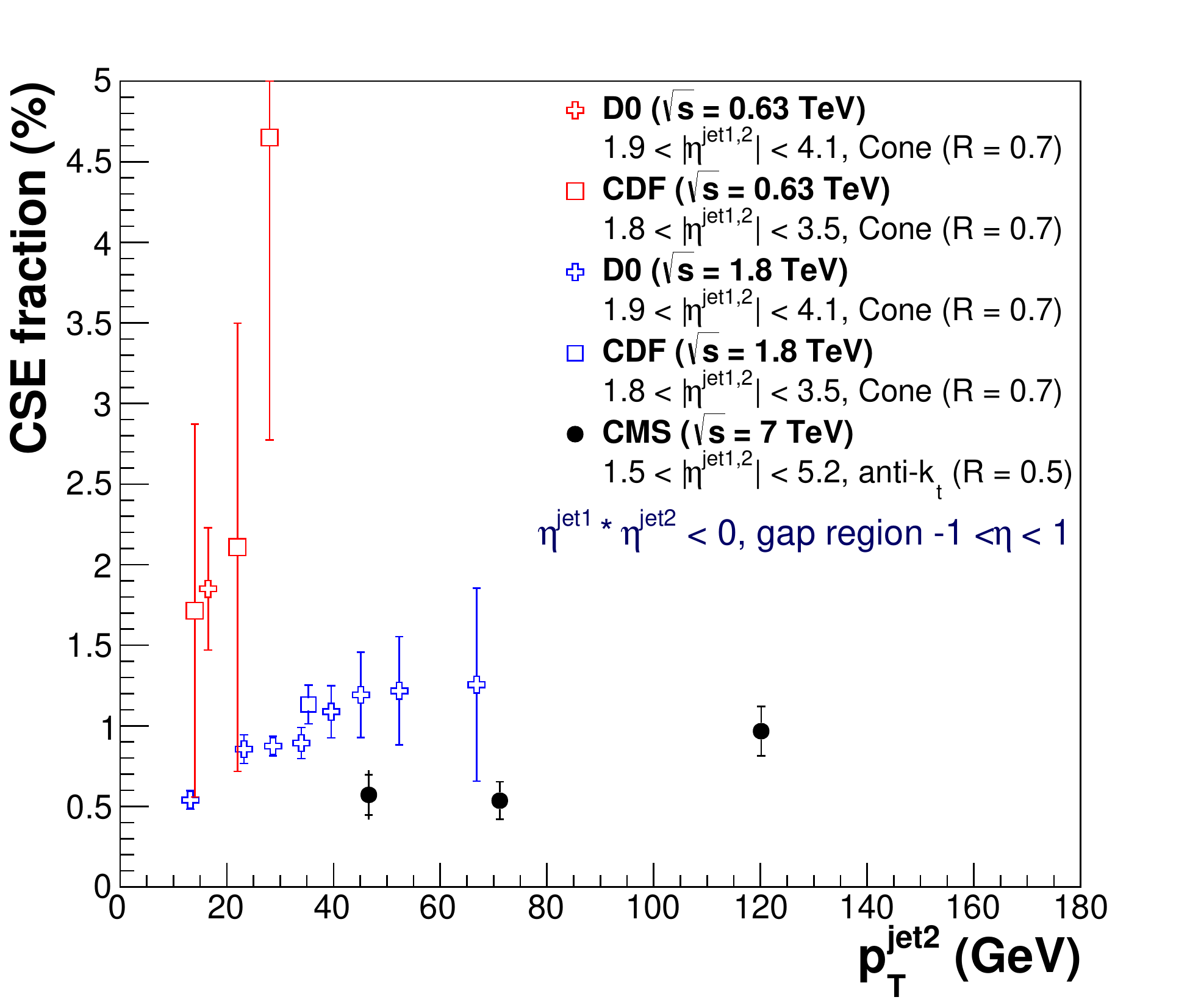}
\caption{Fraction of dijet events with a central gap ($f_\mathrm{CSE}$) as a function of $\pt^\text{jet2}$ at $\sqrt{s}=7$\TeV, compared to the D0~\cite{d03} and CDF~\cite{cdf2,cdf3} results at $\sqrt{s}=0.63$ and 1.8\TeV. The details of the jet selections are given in the legend. The results are plotted at the mean value of $\pt^\text{jet2}$ in the bin. The inner and outer error bars represent the statistical, and the statistical and systematic uncertainties added in quadrature, respectively.}
\label{d0cms}
\end{figure}

The values of the $f_\mathrm{CSE}$ fraction, measured as explained in Section~\ref{sec:jgj} in three bins of $\pt^\text{jet2}$, are given in Table~\ref{pttab}.  Figure~\ref{d0cms} presents the extracted $f_\mathrm{CSE}$ values as a function of $\pt^\text{jet2}$, compared to the results of the D0~\cite{d03} and CDF~\cite{cdf2,cdf3} experiments obtained in similar $\Pp\PAp$ analyses at $\sqrt{s}=0.63$ and 1.8 \TeV. All the measurements are based on the same
pseudorapidity range for the gap region, but differ in the selection of jets. D0 and CDF use the cone jet reconstruction algorithm with size parameter $R = 0.7$, and select jets in the regions $1.9<\abs{\eta^{\text{jet}}}<4.1$, and $1.8<\abs{\eta^{\text{jet}}}<3.5$, respectively. The latter difference only minimally affects the comparison with the CMS results, as the measured $f_\mathrm{CSE}$ fractions at 0.63 and 1.8\TeV depend only weakly on the gap size. At all the three collision energies $f_\mathrm{CSE}$ increases with $\pt^\text{jet2}$. This reflects the fact that the cross section for dijet events with a gap decreases with $\pt^\text{jet2}$ less rapidly than the inclusive dijet cross section does. In addition, a decrease of the gap fraction with increasing $\sqrt{s}$ is observed. The value of $f_\mathrm{CSE}$ measured for $40<\pt^\text{jet2}<60$\GeV at $\sqrt{s}=7$\TeV is about a factor of two lower than those measured for the same $\pt^\text{jet2}$ at $\sqrt{s}=1.8$\TeV. This behavior is in agreement with observations by D0 and CDF, which reported that the jet-gap-jet fraction decreases by a factor of $2.5 \pm 0.9$~\cite{d03} and $3.4 \pm 1.2$~\cite{cdf3}, respectively, when $\sqrt{s}$ increases from 0.63 to 1.8\TeV. The decrease of $f_\mathrm{CSE}$ with increasing energy can be ascribed to a stronger contribution from rescattering processes, in which the interactions between spectator partons destroy the rapidity gap~\cite{bj,glm}. As a consequence, the gap survival probability factor $\abs{S}^2$ is expected to decrease with collision energy. Although no explicit predictions for $\abs{S}^2$ currently exist for jet-gap-jet production at $\sqrt{s}=7$\TeV, a suppression factor of about 2, for $\sqrt{s}$ increasing from 1.8 to 7\TeV, is predicted for central exclusive production~\cite{Khoze:2013dha,Gotsman:2015aga}.

\begin{figure}
\centering
\includegraphics[width=\cmsFigWidth]{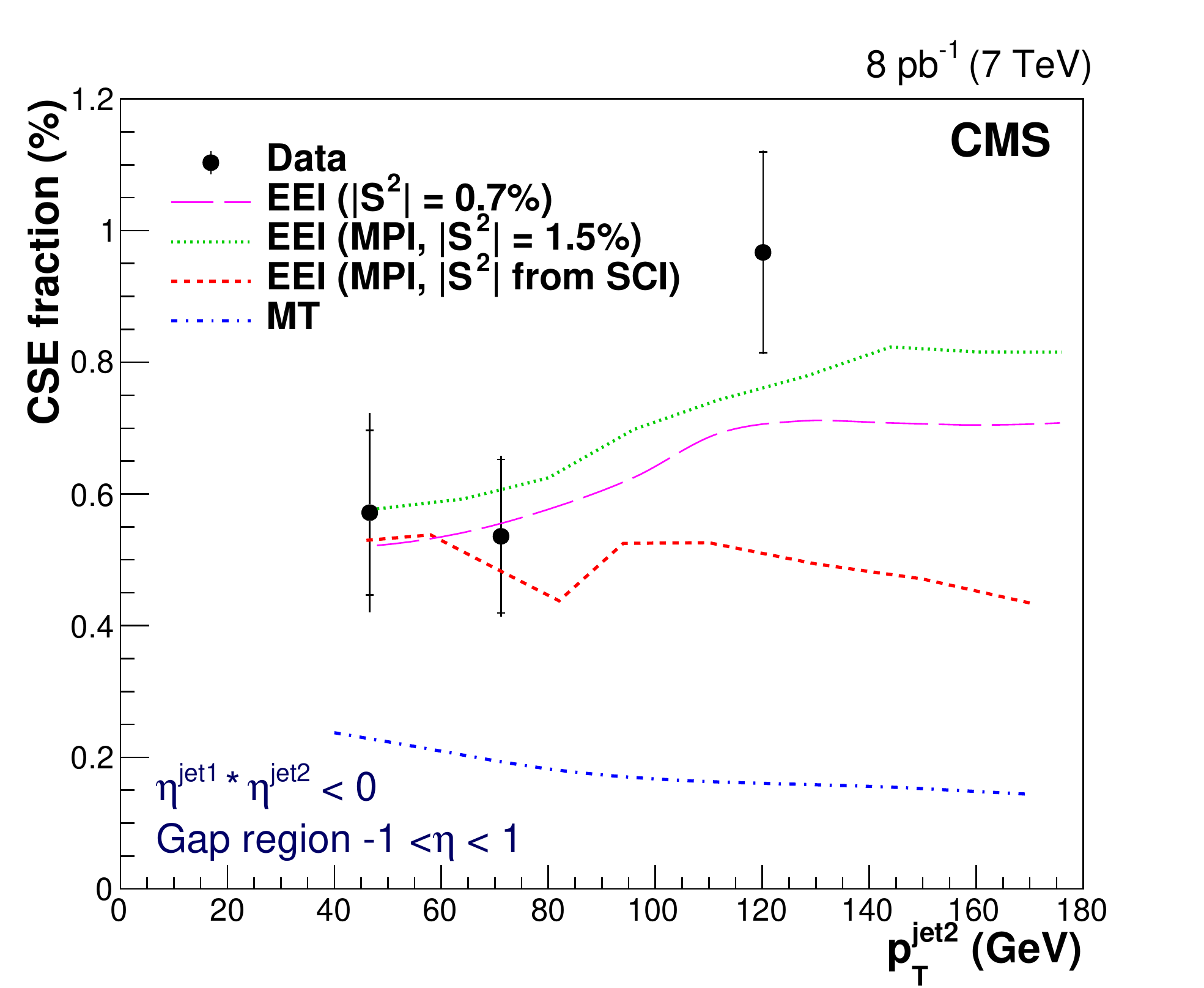}
\caption{Fraction of dijet events with a central gap ($f_\mathrm{CSE}$) as a function of $\pt^\text{jet2}$ at $\sqrt{s}=7$\TeV, compared to the predictions of the Mueller and Tang (MT) model~\cite{mt}, and of the Ekstedt, Enberg, and Ingelman (EEI) model~\cite{csp,cspLHC} with three different treatments of the gap survival probability factor $\abs{S}^2$, as described in the text. The results are plotted at the mean value of $\pt^\text{jet2}$ in the bin. The inner and outer error bars represent the statistical, and the statistical and systematic uncertainties added in quadrature, respectively.}
\label{cmstheory}
\end{figure}

Figure~\ref{cmstheory} shows the comparison of the present results with the BFKL-based theoretical calculations of the Mueller and Tang (MT), and Ekstedt, Enberg and Ingelman (EEI) models. The gap fractions are plotted relative to the standard LO QCD dijet production rates, calculated with \PYTHIA6 (using tune Z2* for MT, and the default settings with color reconnection features turned off for EEI). The MT model~\cite{mt} prediction is based on the LL BFKL evolution in the asymptotic limit of large rapidity separations between the jets, and is obtained with \HERWIG6 (as described in Section~\ref{sec:mc}, without reweighting of the $\pt^\text{jet2}$ dependence) for pure jet-gap-jet events (no simulation of MPI). The MT prediction does not reproduce the increase of $f_\mathrm{CSE}$ with $\pt^\text{jet2}$, as already observed for the 1.8\TeV data~\cite{csp}; it also underestimates the $f_\mathrm{CSE}$ fractions measured at 7\TeV. The EEI predictions~\cite{cspLHC} are based on the model of Ref.~\cite{csp} extended to the present energy. The model includes the dominant next-to-LL corrections to the BFKL evolution of the parton-level cross section, as well as the effect of rescattering processes. For the latter, three approaches are considered, in which gap survival probability is either assumed to be a constant factor, or is partially or fully simulated using Monte Carlo models, to take into account its dependence on the variables $\pt^\text{jet2}$ and $\Delta\eta_\mathrm{jj}$. In the first approach, the BFKL cross section is scaled by a constant factor corresponding to a gap survival probability value of $\abs{S}^2=0.7\%$ (magenta long-dashed curve in Fig.~\ref{cmstheory}), in order to match the data. Alternatively, the activity originating from perturbative gluons is modeled in terms of initial- and final-state parton showers, MPI and hadronization processes, as implemented in \PYTHIA6. The remaining nonperturbative interactions are simulated either by an additional gap survival probability factor of $\abs{S}^2=1.5\%$ (green dotted line in Fig.~\ref{cmstheory}), or by soft color interactions (SCI, red dashed line in Fig.~\ref{cmstheory}) where a color exchange with negligible momentum transfer occurs between parton clusters~\cite{cspLHC}.

As can be seen in Fig.~\ref{cmstheory}, the EEI model with $\abs{S}^2=0.7\%$, and that with MPI and $\abs{S}^2=1.5\%$ reproduce the $\pt^\text{jet2}$ dependence of the $f_\mathrm{CSE}$ fraction in the data. The EEI model with MPI and SCI correctly predicts the amount of jet-gap-jet events in the first two $\pt^\text{jet2}$ bins, but tends to be lower than the data at higher $\pt^\text{jet2}$. The dip in the prediction around $\pt^\text{jet2}=80$\GeV is a result of using the SCI model in conjunction with final state showering, and is a feature of the model rather than a statistical fluctuation.

\begin{table*}
\centering
 \topcaption{Measured values of the fraction of dijet events with a central gap ($f_\mathrm{CSE}$) as a function of the pseudorapidity separation between the jets ($\Delta\eta_\mathrm{jj}$) in bins of $\pt^\text{jet2}$. The columns in the table correspond to $\pt^\text{jet2}$ bins and the rows to $\Delta\eta_\mathrm{jj}$ bins. The first and second errors correspond to the statistical and systematic uncertainties, respectively. The mean values of $\Delta\eta_\mathrm{jj}$ in the bin are also given.}
\label{detatab}
\begin{tabular}{c{c}@{\hspace*{5pt}}cc{c}@{\hspace*{5pt}}cc{c}@{\hspace*{5pt}}cc}
\hline
\pmrule $\pt^\text{jet2}$ (\GeVns{}) && \multicolumn{2}{ c}{ 40--60} &&  \multicolumn{2}{ c }{ 60--100} &&  \multicolumn{2}{c }{ 100--200} \\ \cline{1-1}\cline{3-4} \cline{6-7} \cline{9-10}
\pmrule $\Delta\eta_\mathrm{jj}$ range && $\langle \Delta\eta_\mathrm{jj} \rangle$ & $f_\mathrm{CSE}$ (\%) && $\langle \Delta\eta_\mathrm{jj} \rangle$ & $f_\mathrm{CSE}$ (\%) && $\langle \Delta\eta_\mathrm{jj} \rangle$  & $f_\mathrm{CSE}$ (\%)\\
\hline
\pmrule 3--4 && 3.63 & $0.25 \pm 0.20 \pm 0.04$  &&  3.62 & $0.47 \pm 0.19 \pm 0.05$  && 3.61 & $0.78 \pm 0.21 \pm 0.06$ \\
\pmrule 4--5 && 4.46 & $0.41 \pm 0.16 \pm 0.14$  &&  4.45 & $0.47 \pm 0.16 \pm 0.08$  && 4.41 & $0.99 \pm 0.23 \pm 0.06$ \\
\pmrule 5--7 && 5.60 & $1.24 \pm 0.32 \pm 0.10$  &&  5.49 & $0.91 \pm 0.32 \pm 0.21$  && 5.37 & $1.95 \pm 0.69 \pm 0.44$ \\
\hline
\end{tabular}
\end{table*}

\begin{figure} \centering
\includegraphics[width=0.45\textwidth]{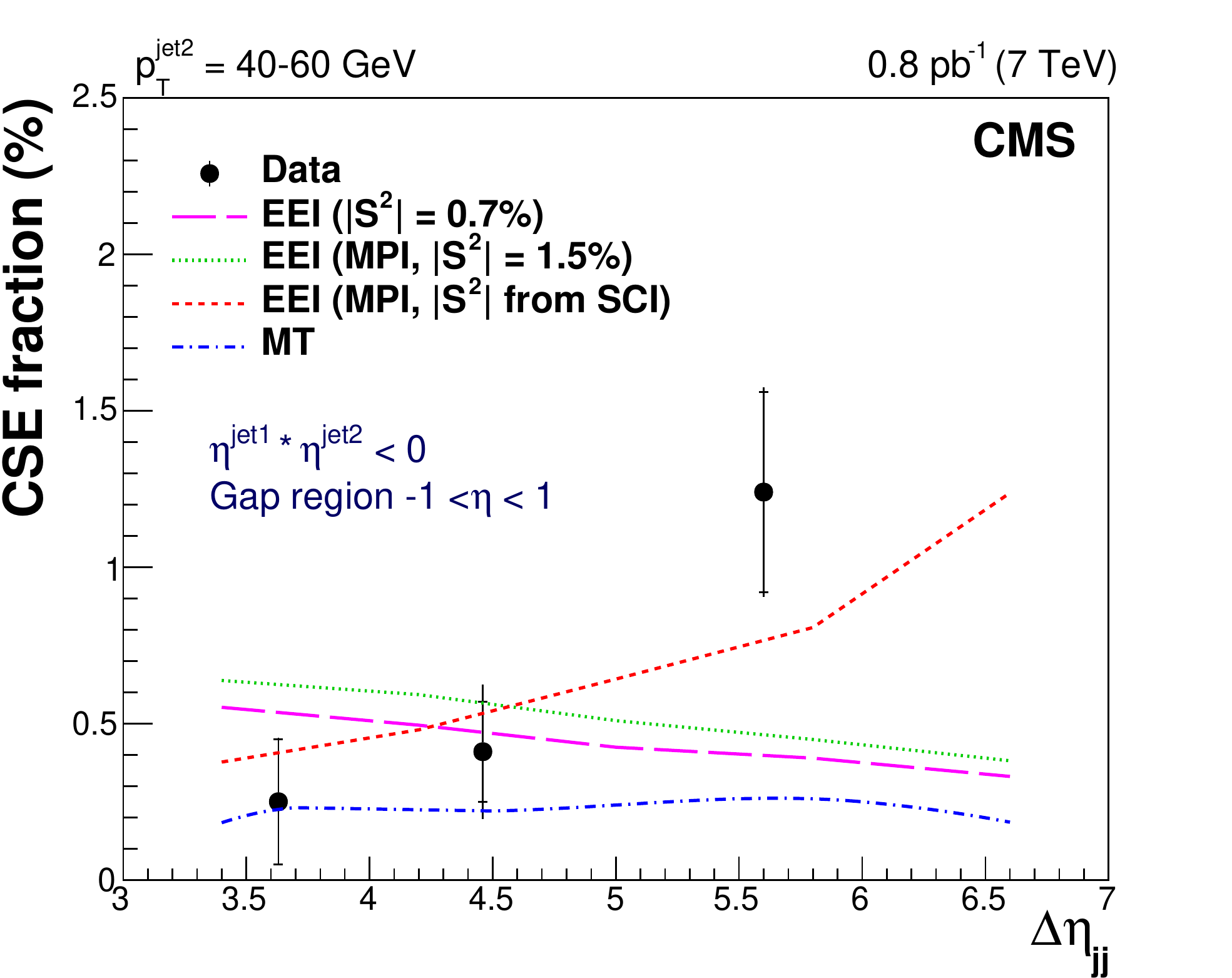}
\includegraphics[width=0.45\textwidth]{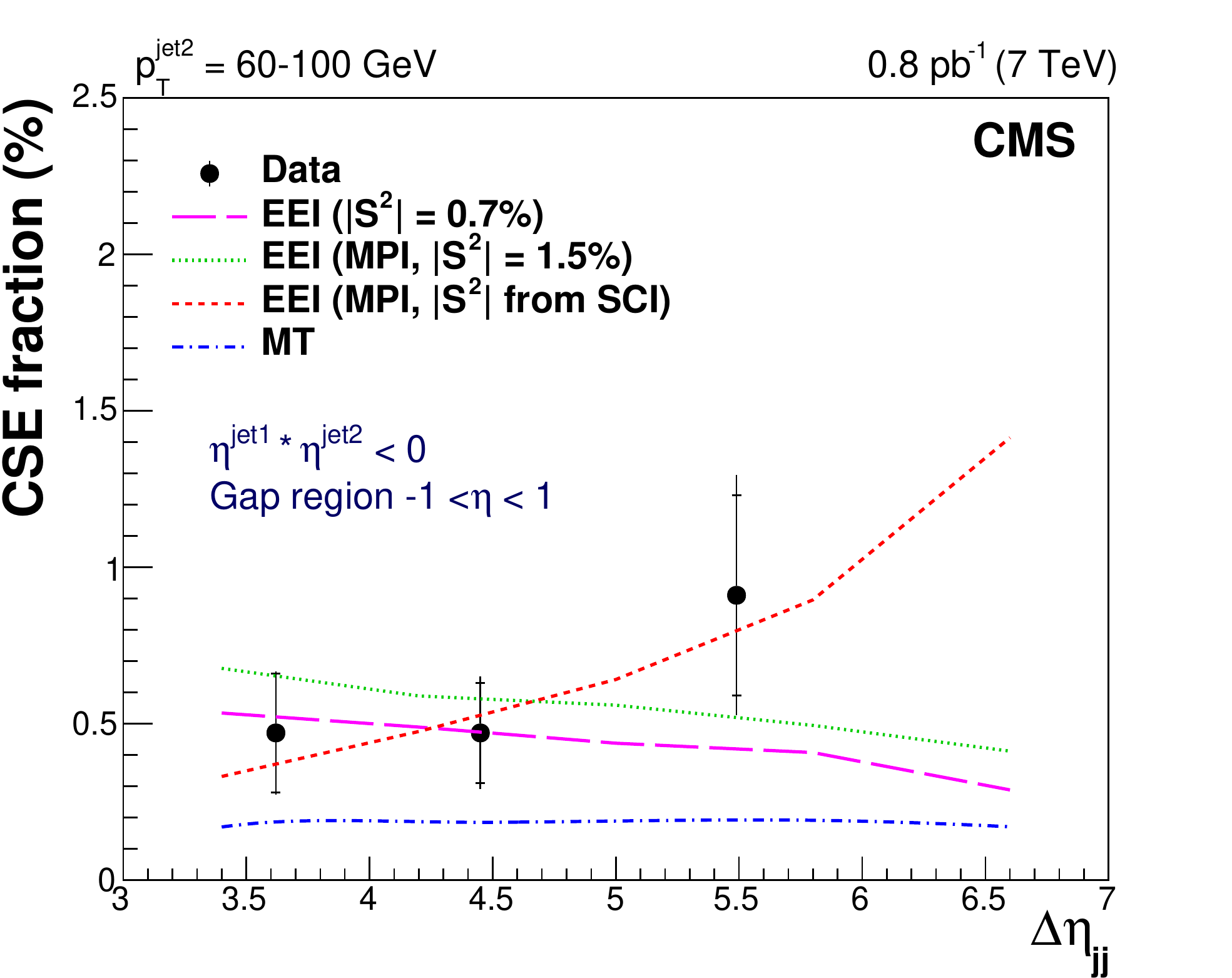}
\includegraphics[width=0.45\textwidth]{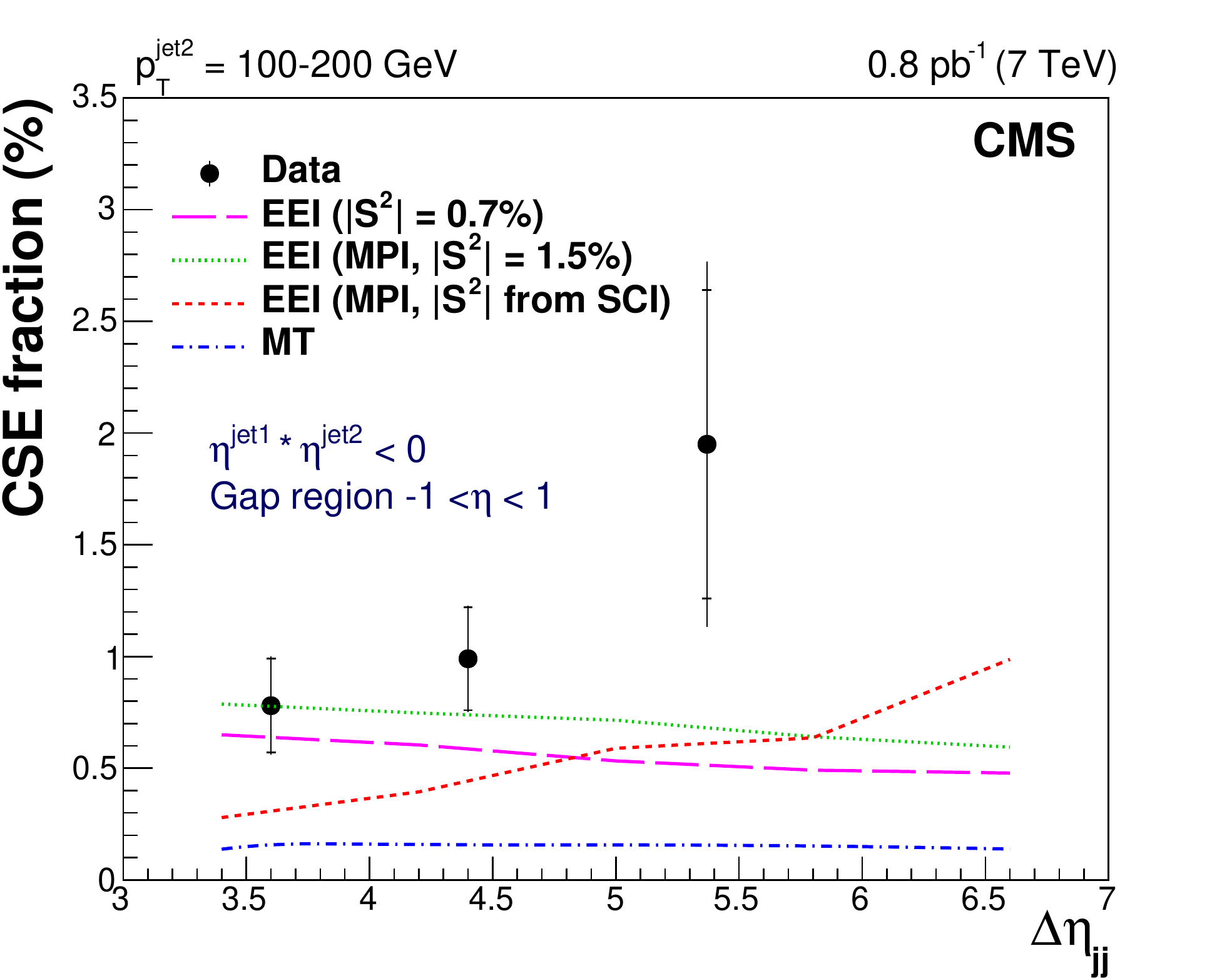}
\caption{Fraction of dijet events with a central gap ($f_\mathrm{CSE}$) as a function of $\Delta\eta_\mathrm{jj}$ at $\sqrt{s}=7$\TeV in three different $\pt^\text{jet2}$ ranges, compared to the predictions of the Mueller and Tang (MT) model~\cite{mt}, and of the Ekstedt, Enberg, and Ingelman (EEI) model~\cite{csp,cspLHC} with three different treatments of the gap survival probability factor $\abs{S}^2$, as described in the text. The results are plotted at the mean value of $\Delta\eta_\mathrm{jj}$ in the bin. Inner and outer error bars correspond to the statistical, and the statistical and systematic uncertainties added in quadrature, respectively.}
\label{fineta}
\end{figure}

The dependence of the $f_\mathrm{CSE}$ fraction on the size of $\Delta\eta_\mathrm{jj}$ is studied for each $\pt^\text{jet2}$ sample in three bins of $\Delta\eta_\mathrm{jj}$ = 3--4, 4--5, and 5--7. The measured values of the $f_\mathrm{CSE}$ fractions are listed in Table~\ref{detatab}, and plotted in Fig.~\ref{fineta}. The fraction of jet-gap-jet events increases with $\Delta\eta_\mathrm{jj}$, and varies from 0.3 to 1.2\%, and from 0.8 to 2\%, in the lowest and the highest $\pt^\text{jet2}$ bins, respectively. Figure~\ref{fineta} also shows the comparison of the data with the predictions of the MT and EEI models. The MT model predicts a flat dependence of $f_\mathrm{CSE}$ with $\Delta\eta_\mathrm{jj}$, and underestimates the measured jet-gap-jet fractions except for the lowest ($\pt^\text{jet2}$, $\Delta\eta_\mathrm{jj}$) bin for which the agreement is good. The EEI model with the $\abs{S}^2=0.7\%$ factor, as well as that with MPI plus $\abs{S}^2=1.5\%$ predict a decrease of $f_\mathrm{CSE}$ with $\Delta\eta_\mathrm{jj}$, and are at variance with the data. Conversely, the EEI model with MPI plus soft color interactions satisfactorily reproduces the rise of $f_\mathrm{CSE}$ with $\Delta\eta_\mathrm{jj}$ in all $\pt^\text{jet2}$ bins.

\section{Summary}

Events with a large rapidity gap between the two leading jets have been measured for the first time at the LHC, for jets with transverse momentum $\pt^{\text{jet}}>40$\GeV and pseudorapidity $1.5<\abs{\eta^{\text{jet}}}<4.7$, reconstructed in opposite ends of the detector. The number of dijet events with no particles with $\pt>0.2$\GeV in the region $\abs{\eta}<1$ is severely underestimated by \PYTHIA6 (tune Z2*). \HERWIG6 predictions, which include a contribution from color singlet exchange (CSE), based on the leading logarithmic Balitsky--Fadin--Kuraev--Lipatov (BFKL) evolution equations, are needed to reproduce the type of dijet topologies selected in our analysis. The fraction of selected dijet events with  such a rapidity gap has been measured as a function of the second-leading jet transverse momentum ($\pt^\text{jet2}$) and as a function of the size of the pseudorapidity interval between the jets, $\Delta\eta_\mathrm{jj}$. The $f_\mathrm{CSE}$  fraction rises with $\pt^\text{jet2}$ (from 0.6 to 1\%) and with $\Delta\eta_\mathrm{jj}$ (from 0.3 to 1.2\% for $40<\pt^\text{jet2}<60$\GeV, from 0.5 to 0.9\% for $60<\pt^\text{jet2}<100$\GeV, and from 0.8 to 2\% for $100<\pt^\text{jet2}<200$\GeV).

The measured CSE fractions have been compared to the results of the D0 and CDF experiments at a center-of-mass energies of 0.63 and 1.8\TeV. A factor of two decrease of the CSE fraction measured at $\sqrt{s} = 7$\TeV with respect to that at $\sqrt{s} = 1.8$\TeV is observed. Such a behavior is consistent with the decrease seen in the Tevatron data when $\sqrt{s}$ rises from 0.63 to~1.8\TeV, and with theoretical expectations for the $\sqrt{s}$ dependence of the rapidity gap survival probability.

The data are also compared to theoretical perturbative quantum chromodynamics calculations based on the BFKL evolution equations complemented with different estimates of the non-perturbative gap survival probability. The next-to-leading-logarithmic BFKL calculations of Ekstedt, Enberg and Ingelman, with three different implementations of the soft rescattering processes, describe many features of the data, but none of the implementations is able to simultaneously describe all the features of the measurement.

\begin{acknowledgments} We would like to thank Andreas Ekstedt, Rikard Enberg, and Gunnar Ingelman for providing the next-to-LL BFKL analytical predictions of their model.

We congratulate our colleagues in the CERN accelerator departments for the excellent performance of the LHC and thank the technical and administrative staffs at CERN and at other CMS institutes for their contributions to the success of the CMS effort. In addition, we gratefully acknowledge the computing centers and personnel of the Worldwide LHC Computing Grid for delivering so effectively the computing infrastructure essential to our analyses. Finally, we acknowledge the enduring support for the construction and operation of the LHC and the CMS detector provided by the following funding agencies: BMWFW and FWF (Austria); FNRS and FWO (Belgium); CNPq, CAPES, FAPERJ, and FAPESP (Brazil); MES (Bulgaria); CERN; CAS, MoST, and NSFC (China); COLCIENCIAS (Colombia); MSES and CSF (Croatia); RPF (Cyprus); SENESCYT (Ecuador); MoER, ERC IUT, and ERDF (Estonia); Academy of Finland, MEC, and HIP (Finland); CEA and CNRS/IN2P3 (France); BMBF, DFG, and HGF (Germany); GSRT (Greece); OTKA and NIH (Hungary); DAE and DST (India); IPM (Iran); SFI (Ireland); INFN (Italy); MSIP and NRF (Republic of Korea); LAS (Lithuania); MOE and UM (Malaysia); BUAP, CINVESTAV, CONACYT, LNS, SEP, and UASLP-FAI (Mexico); MBIE (New Zealand); PAEC (Pakistan); MSHE and NSC (Poland); FCT (Portugal); JINR (Dubna); MON, RosAtom, RAS, RFBR and RAEP (Russia); MESTD (Serbia); SEIDI, CPAN, PCTI and FEDER (Spain); Swiss Funding Agencies (Switzerland); MST (Taipei); ThEPCenter, IPST, STAR, and NSTDA (Thailand); TUBITAK and TAEK (Turkey); NASU and SFFR (Ukraine); STFC (United Kingdom); DOE and NSF (USA).

\hyphenation{Rachada-pisek} Individuals have received support from the Marie-Curie program and the European Research Council and Horizon 2020 Grant, contract No. 675440 (European Union); the Leventis Foundation; the A. P. Sloan Foundation; the Alexander von Humboldt Foundation; the Belgian Federal Science Policy Office; the Fonds pour la Formation \`a la Recherche dans l'Industrie et dans l'Agriculture (FRIA-Belgium); the Agentschap voor Innovatie door Wetenschap en Technologie (IWT-Belgium); the Ministry of Education, Youth and Sports (MEYS) of the Czech Republic; the Council of Science and Industrial Research, India; the HOMING PLUS program of the Foundation for Polish Science, cofinanced from European Union, Regional Development Fund, the Mobility Plus program of the Ministry of Science and Higher Education, the National Science Center (Poland), contracts Harmonia 2014/14/M/ST2/00428, Opus 2014/13/B/ST2/02543, 2014/15/B/ST2/03998, and 2015/19/B/ST2/02861, Sonata-bis 2012/07/E/ST2/01406; the National Priorities Research Program by Qatar National Research Fund; the Programa Severo Ochoa del Principado de Asturias; the Thalis and Aristeia programs cofinanced by EU-ESF and the Greek NSRF; the Rachadapisek Sompot Fund for Postdoctoral Fellowship, Chulalongkorn University and the Chulalongkorn Academic into Its 2nd Century Project Advancement Project (Thailand); the Welch Foundation, contract C-1845; and the Weston Havens Foundation (USA).
\end{acknowledgments}

\bibliography{auto_generated}

\providecommand{\href}[2]{#2}\begingroup\raggedright\begin{thebibliography}{10}%
\makeatletter
\providecommand{\hrefCMSnoop }[0]{\@secondoftwo}%
\makeatother
\providecommand{\doi}{\texttt{doi:}\begingroup \urlstyle{tt}\Url}

\bibitem{Khachatryan:2011zj}
\hrefCMSnoop {}{{CMS Collaboration}, ``Dijet azimuthal decorrelations in $pp$
  collisions at {$\sqrt{s} = 7$~TeV}'',} \textit{ Phys. Rev. Lett.} \textbf{
  106} (2011) 122003,
  \href{http://dx.doi.org/10.1103/PhysRevLett.106.122003}{\doi{10.1103/PhysRevLett.106.122003}},
\href{http://www.arXiv.org/abs/1101.5029}{\texttt{arXiv:1101.5029}}.

\bibitem{Chatrchyan:2011qta}
\hrefCMSnoop {}{{CMS Collaboration}, ``Measurement of the differential dijet
  production cross section in proton-proton collisions at {$\sqrt{s}=7$
  TeV}'',} \textit{ Phys. Lett. B} \textbf{ 700} (2011) 187,
  \href{http://dx.doi.org/10.1016/j.physletb.2011.05.027}{\doi{10.1016/j.physletb.2011.05.027}},
\href{http://www.arXiv.org/abs/1104.1693}{\texttt{arXiv:1104.1693}}.

\bibitem{Chatrchyan:2012gwa}
\hrefCMSnoop {}{{CMS Collaboration}, ``{Measurement of the inclusive production
  cross sections for forward jets and for dijet events with one forward and one
  central jet in pp collisions at $\sqrt{s}=7$ TeV}'',} \textit{ JHEP} \textbf{
  06} (2012) 036,
  \href{http://dx.doi.org/10.1007/JHEP06(2012)036}{\doi{10.1007/JHEP06(2012)036}},
\href{http://www.arXiv.org/abs/1202.0704}{\texttt{arXiv:1202.0704}}.

\bibitem{Chatrchyan:2012pb}
\hrefCMSnoop {}{{CMS Collaboration}, ``Ratios of dijet production cross
  sections as a function of the absolute difference in rapidity between jets in
  proton-proton collisions at {$\sqrt{s}=7$ TeV}'',} \textit{ Eur. Phys. J. C}
  \textbf{ 72} (2012) 2216,
  \href{http://dx.doi.org/10.1140/epjc/s10052-012-2216-6}{\doi{10.1140/epjc/s10052-012-2216-6}},
\href{http://www.arXiv.org/abs/1204.0696}{\texttt{arXiv:1204.0696}}.

\bibitem{Khachatryan:2016hkr}
\hrefCMSnoop {}{{CMS Collaboration}, ``Measurement of dijet azimuthal
  decorrelation in pp collisions at {$\sqrt{s}=8$ TeV}'',} \textit{ Eur. Phys.
  J. C} \textbf{ 76} (2016) 536,
  \href{http://dx.doi.org/10.1140/epjc/s10052-016-4346-8}{\doi{10.1140/epjc/s10052-016-4346-8}},
\href{http://www.arXiv.org/abs/1602.04384}{\texttt{arXiv:1602.04384}}.

\bibitem{Sirunyan:2017skj}
\hrefCMSnoop {}{{CMS Collaboration}, ``{Measurement of the triple-differential
  dijet cross section in proton-proton collisions at $\sqrt{s}=8\,\text {TeV} $
  and constraints on parton distribution functions}'',} \textit{ Eur. Phys. J.}
  \textbf{ C77} (2017) 746,
  \href{http://dx.doi.org/10.1140/epjc/s10052-017-5286-7}{\doi{10.1140/epjc/s10052-017-5286-7}},
\href{http://www.arXiv.org/abs/1705.02628}{\texttt{arXiv:1705.02628}}.

\bibitem{Aad:2010ad}
\hrefCMSnoop {}{{ATLAS Collaboration}, ``Measurement of inclusive jet and dijet
  cross sections in proton-proton collisions at {7 TeV} centre-of-mass energy
  with the {ATLAS} detector'',} \textit{ Eur. Phys. J. C} \textbf{ 71} (2011)
  1512,
  \href{http://dx.doi.org/10.1140/epjc/s10052-010-1512-2}{\doi{10.1140/epjc/s10052-010-1512-2}},
\href{http://www.arXiv.org/abs/1009.5908}{\texttt{arXiv:1009.5908}}.

\bibitem{Aad:2011jz}
\hrefCMSnoop {}{{ATLAS Collaboration}, ``Measurement of dijet production with a
  veto on additional central jet activity in pp collisions at {$\sqrt{s}=7$
  TeV} using the {ATLAS} detector'',} \textit{ JHEP} \textbf{ 09} (2011) 053,
  \href{http://dx.doi.org/10.1007/JHEP09(2011)053}{\doi{10.1007/JHEP09(2011)053}},
\href{http://www.arXiv.org/abs/1107.1641}{\texttt{arXiv:1107.1641}}.

\bibitem{Aad:2011fc}
\hrefCMSnoop {}{{ATLAS Collaboration}, ``Measurement of inclusive jet and dijet
  production in pp collisions at {$\sqrt{s}=7$ TeV} using the {ATLAS}
  detector'',} \textit{ Phys. Rev. D} \textbf{ 86} (2012) 014022,
  \href{http://dx.doi.org/10.1103/PhysRevD.86.014022}{\doi{10.1103/PhysRevD.86.014022}},
\href{http://www.arXiv.org/abs/1112.6297}{\texttt{arXiv:1112.6297}}.

\bibitem{Aad:2013tea}
\hrefCMSnoop {}{{ATLAS Collaboration}, ``Measurement of dijet cross sections in
  pp collisions at {7 TeV} centre-of-mass energy using the {ATLAS} detector'',}
  \textit{ JHEP} \textbf{ 05} (2014) 059,
  \href{http://dx.doi.org/10.1007/JHEP05(2014)059}{\doi{10.1007/JHEP05(2014)059}},
\href{http://www.arXiv.org/abs/1312.3524}{\texttt{arXiv:1312.3524}}.

\bibitem{Aad:2014pua}
\hrefCMSnoop {}{{ATLAS Collaboration}, ``Measurements of jet vetoes and
  azimuthal decorrelations in dijet events produced in pp collisions at
  {$\sqrt{s}=7$ TeV} using the {ATLAS} detector'',} \textit{ Eur. Phys. J. C}
  \textbf{ 74} (2014) 3117,
  \href{http://dx.doi.org/10.1140/epjc/s10052-014-3117-7}{\doi{10.1140/epjc/s10052-014-3117-7}},
\href{http://www.arXiv.org/abs/1407.5756}{\texttt{arXiv:1407.5756}}.

\bibitem{Aad:2015xis}
\hrefCMSnoop {}{{ATLAS Collaboration}, ``Dijet production in {$\sqrt{s}=$ 7
  TeV} pp collisions with large rapidity gaps at the {ATLAS} experiment'',}
  \textit{ Phys. Lett. B} \textbf{ 754} (2016) 214,
  \href{http://dx.doi.org/10.1016/j.physletb.2016.01.028}{\doi{10.1016/j.physletb.2016.01.028}},
\href{http://www.arXiv.org/abs/1511.00502}{\texttt{arXiv:1511.00502}}.

\bibitem{dglap1}
\hrefCMSnoop {}{V.~N. Gribov and L.~N. Lipatov, ``Deep inelastic ep scattering
  in perturbation theory'',} \textit{ Sov. J. Nucl. Phys.} \textbf{ 15} (1972)
438.

\bibitem{dglap2}
\hrefCMSnoop {}{G.~Altarelli and G.~Parisi, ``Asymptotic freedom in parton
  language'',} \textit{ Nucl. Phys. B} \textbf{ 126} (1977) 298,
\href{http://dx.doi.org/10.1016/0550-3213(77)90384-4}{\doi{10.1016/0550-3213(77)90384-4}}.

\bibitem{dglap3}
\hrefCMSnoop {}{Y.~L. Dokshitzer, ``Calculation of the structure functions for
  deep inelastic scattering and e+ e- annihilation by perturbation theory in
  quantum chromodynamics'',} \textit{ Sov. Phys. JETP} \textbf{ 46} (1977)
641.

\bibitem{bfkl1}
\hrefCMSnoop {}{E.~E. Kuraev, L.~N. Lipatov, and V.~S. Fadin, ``The
  {P}omeranchuk singularity in nonabelian gauge theories'',} \textit{ Sov.
  Phys. JETP} \textbf{ 45} (1977) 199.

\bibitem{bfkl2}
\hrefCMSnoop {}{Y.~Y. Balitsky and L.~N. Lipatov, ``The {P}omeranchuk
  singularity in quantum chromodynamics'',} \textit{ Sov. J. Nucl. Phys.}
  \textbf{ 28} (1978) 822.

\bibitem{bfkl3}
\hrefCMSnoop {}{L.~N. Lipatov, ``The bare pomeron in quantum chromodynamics'',}
  \textit{ Sov. Phys. JETP} \textbf{ 63} (1986) 904.

\bibitem{bj}
\hrefCMSnoop {}{J.~D. Bjorken, ``Rapidity gaps and jets as a new physics
  signature in very high-energy hadron hadron collisions'',} \textit{ Phys.
  Rev. D} \textbf{ 47} (1993) 101,
\href{http://dx.doi.org/10.1103/PhysRevD.47.101}{\doi{10.1103/PhysRevD.47.101}}.

\bibitem{diffraction}
V.~Barone and E.~Predazzi, ``High-energy particle diffraction''.
\newblock Texts and monographs in physics. Springer, Berlin, 2002.
\newblock
  \href{http://dx.doi.org/10.1007/978-3-662-04724-8}{\doi{10.1007/978-3-662-04724-8}}.

\bibitem{mt}
\hrefCMSnoop {}{A.~H. Mueller and W.-K. Tang, ``{High-energy parton-parton
  elastic scattering in QCD}'',} \textit{ Phys. Lett. B} \textbf{ 284} (1992)
  123,
\href{http://dx.doi.org/10.1016/0370-2693(92)91936-4}{\doi{10.1016/0370-2693(92)91936-4}}.

\bibitem{csp}
\hrefCMSnoop {}{R.~Enberg, G.~Ingelman, and L.~Motyka, ``{Hard color singlet
  exchange and gaps between jets at the Tevatron}'',} \textit{ Phys. Lett. B}
  \textbf{ 524} (2002) 273,
  \href{http://dx.doi.org/10.1016/S0370-2693(01)01379-X}{\doi{10.1016/S0370-2693(01)01379-X}},
\href{http://www.arXiv.org/abs/hep-ph/0111090}{\texttt{arXiv:hep-ph/0111090}}.

\bibitem{cspLHC}
\hrefCMSnoop {}{A.~Ekstedt, R.~Enberg, and G.~Ingelman, ``{Hard color singlet
  BFKL exchange and gaps between jets at the LHC}'',}
\newblock 2017.
\newblock
\href{http://www.arXiv.org/abs/1703.10919}{\texttt{arXiv:1703.10919}}.
\newblock

\bibitem{kmr}
\hrefCMSnoop {}{O.~Kepka, C.~Marquet, and C.~Royon, ``{Gaps between jets in
  hadronic collisions}'',} \textit{ Phys. Rev. D} \textbf{ 83} (2011) 034036,
  \href{http://dx.doi.org/10.1103/PhysRevD.83.034036}{\doi{10.1103/PhysRevD.83.034036}},
\href{http://www.arXiv.org/abs/1012.3849}{\texttt{arXiv:1012.3849}}.

\bibitem{d01}
\hrefCMSnoop {}{{D0} Collaboration, ``{Rapidity gaps between jets in
  p$\bar{\mathrm{p}}$ collisions at $\sqrt{s} = 1.8$ TeV}'',} \textit{ Phys.
  Rev. Lett.} \textbf{ 72} (1994) 2332,
\href{http://dx.doi.org/10.1103/PhysRevLett.72.2332}{\doi{10.1103/PhysRevLett.72.2332}}.

\bibitem{d02}
\hrefCMSnoop {}{{D0} Collaboration, ``Jet production via strongly-interacting
  color-singlet exchange in p$\bar{\mathrm{p}}$ collisions'',} \textit{ Phys.
  Rev. Lett.} \textbf{ 76} (1996) 734,
  \href{http://dx.doi.org/10.1103/PhysRevLett.76.734}{\doi{10.1103/PhysRevLett.76.734}},
\href{http://www.arXiv.org/abs/hep-ex/9509013}{\texttt{arXiv:hep-ex/9509013}}.

\bibitem{d03}
\hrefCMSnoop {}{{D0} Collaboration, ``{Probing hard color-singlet exchange in
  p$\bar{\mathrm{p}}$ collisions at $\sqrt{s} = 630$ GeV and 1800 GeV}'',}
  \textit{ Phys. Lett. B} \textbf{ 440} (1998) 189,
  \href{http://dx.doi.org/10.1016/S0370-2693(98)01238-6}{\doi{10.1016/S0370-2693(98)01238-6}},
\href{http://www.arXiv.org/abs/hep-ex/9809016}{\texttt{arXiv:hep-ex/9809016}}.

\bibitem{cdf1}
\hrefCMSnoop {}{{CDF} Collaboration, ``{Observation of rapidity gaps in
  $\bar{\mathrm{p}}$p collisions at 1.8 TeV}'',} \textit{ Phys. Rev. Lett.}
  \textbf{ 74} (1995) 855,
\href{http://dx.doi.org/10.1103/PhysRevLett.74.855}{\doi{10.1103/PhysRevLett.74.855}}.

\bibitem{cdf2}
\hrefCMSnoop {}{{CDF} Collaboration, ``{Dijet production by color-singlet
  exchange at the Fermilab Tevatron}'',} \textit{ Phys. Rev. Lett.} \textbf{
  80} (1998) 1156,
\href{http://dx.doi.org/10.1103/PhysRevLett.80.1156}{\doi{10.1103/PhysRevLett.80.1156}}.

\bibitem{cdf3}
\hrefCMSnoop {}{{CDF} Collaboration, ``{Events with a rapidity gap between jets
  in $\bar{\mathrm{p}}$p collisions at $\sqrt{s} = 630$ GeV}'',} \textit{ Phys.
  Rev. Lett.} \textbf{ 81} (1998) 5278,
\href{http://dx.doi.org/10.1103/PhysRevLett.81.5278}{\doi{10.1103/PhysRevLett.81.5278}}.

\bibitem{zeus}
\hrefCMSnoop {}{{ZEUS} Collaboration, ``{Rapidity gaps between jets in
  photoproduction at HERA}'',} \textit{ Phys. Lett. B} \textbf{ 369} (1996) 55,
  \href{http://dx.doi.org/10.1016/0370-2693(95)01588-4}{\doi{10.1016/0370-2693(95)01588-4}},
\href{http://www.arXiv.org/abs/hep-ex/9510012}{\texttt{arXiv:hep-ex/9510012}}.

\bibitem{h1}
\hrefCMSnoop {}{{H1} Collaboration, ``{Energy flow and rapidity gaps between
  jets in photoproduction at HERA}'',} \textit{ Eur. Phys. J. C} \textbf{ 24}
  (2002) 517,
  \href{http://dx.doi.org/10.1007/s10052-002-0988-9}{\doi{10.1007/s10052-002-0988-9}},
\href{http://www.arXiv.org/abs/hep-ex/0203011}{\texttt{arXiv:hep-ex/0203011}}.

\bibitem{TRK-11-001}
\hrefCMSnoop {}{{CMS Collaboration}, ``{Description and performance of track
  and primary-vertex reconstruction with the CMS tracker}'',} \textit{ JINST}
  \textbf{ 9} (2014) P10009,
  \href{http://dx.doi.org/10.1088/1748-0221/9/10/P10009}{\doi{10.1088/1748-0221/9/10/P10009}},
\href{http://www.arXiv.org/abs/1405.6569}{\texttt{arXiv:1405.6569}}.

\bibitem{Chatrchyan:2008zzk}
\hrefCMSnoop {}{{CMS Collaboration}, ``The {CMS} experiment at the {CERN}
  {LHC}'',} \textit{ JINST} \textbf{ 3} (2008) S08004,
  \href{http://dx.doi.org/10.1088/1748-0221/3/08/S08004}{\doi{10.1088/1748-0221/3/08/S08004}}.

\bibitem{CMStrigger}
\hrefCMSnoop {}{{CMS Collaboration}, ``{The CMS trigger system}'',} \textit{
  JINST} \textbf{ 12} (2017) P01020,
  \href{http://dx.doi.org/10.1088/1748-0221/12/01/P01020}{\doi{10.1088/1748-0221/12/01/P01020}},
\href{http://www.arXiv.org/abs/1609.02366}{\texttt{arXiv:1609.02366}}.

\bibitem{Cacciari:2008gp}
\hrefCMSnoop {}{M.~Cacciari, G.~P. Salam, and G.~Soyez, ``{The anti-$k_t$ jet
  clustering algorithm}'',} \textit{ JHEP} \textbf{ 04} (2008) 063,
  \href{http://dx.doi.org/10.1088/1126-6708/2008/04/063}{\doi{10.1088/1126-6708/2008/04/063}},
  \href{http://www.arXiv.org/abs/0802.1189}{\texttt{arXiv:0802.1189}}.

\bibitem{Cacciari:2011ma}
\hrefCMSnoop {}{M.~Cacciari, G.~P. Salam, and G.~Soyez, ``{FastJet user
  manual}'',} \textit{ Eur. Phys. J. C} \textbf{ 72} (2012) 1896,
  \href{http://dx.doi.org/10.1140/epjc/s10052-012-1896-2}{\doi{10.1140/epjc/s10052-012-1896-2}},
\href{http://www.arXiv.org/abs/1111.6097}{\texttt{arXiv:1111.6097}}.

\bibitem{PFnew}
\hrefCMSnoop {}{{CMS Collaboration}, ``{Particle-flow reconstruction and global
  event description with the CMS detector}'',} \textit{ JINST} \textbf{ 12}
  (2017) P10003,
  \href{http://dx.doi.org/10.1088/1748-0221/12/10/P10003}{\doi{10.1088/1748-0221/12/10/P10003}},
\href{http://www.arXiv.org/abs/1706.04965}{\texttt{arXiv:1706.04965}}.

\bibitem{CMS-PAS-JME-10-003}
\href {https://cds.cern.ch/record/1279362}{{CMS Collaboration}, ``Jet
  performance in pp collisions at {7 TeV}'',} CMS Physics Analysis Summary
  CMS-PAS-JME-10-003, 2010.

\bibitem{CMS-PAS-JME-10-010}
\href {http://cdsweb.cern.ch/record/1308178}{{CMS Collaboration},
  ``Determination of the jet energy scale in {CMS} with pp collisions at
  $\sqrt{s}= 7$ {TeV}'',} CMS Physics Analysis Summary CMS-PAS-JME-10-010,
  2010.

\bibitem{CMS-JME-10-011}
\hrefCMSnoop {}{{CMS Collaboration}, ``Determination of jet energy calibration
  and transverse momentum resolution in {CMS}'',} \textit{ JINST} \textbf{ 6}
  (2011) P11002,
  \href{http://dx.doi.org/10.1088/1748-0221/6/11/P11002}{\doi{10.1088/1748-0221/6/11/P11002}},
\href{http://www.arXiv.org/abs/1107.4277}{\texttt{arXiv:1107.4277}}.

\bibitem{pythia}
\hrefCMSnoop {}{T.~Sj{\"o}strand, S.~Mrenna, and P.~Skands, ``{PYTHIA} 6.4
  physics and manual'',} \textit{ JHEP} \textbf{ 05} (2006) 026,
  \href{http://dx.doi.org/10.1088/1126-6708/2006/05/026}{\doi{10.1088/1126-6708/2006/05/026}},
\href{http://www.arXiv.org/abs/hep-ph/0603175}{\texttt{arXiv:hep-ph/0603175}}.

\bibitem{Andersson:1983ia}
\hrefCMSnoop {}{B.~Andersson, G.~Gustafson, G.~Ingelman, and T.~Sj{\"o}strand,
  ``Parton fragmentation and string dynamics'',} \textit{ Phys. Rept.} \textbf{
  97} (1983) 31,
\href{http://dx.doi.org/10.1016/0370-1573(83)90080-7}{\doi{10.1016/0370-1573(83)90080-7}}.

\bibitem{Chatrchyan:2013gfi}
\hrefCMSnoop {}{{CMS Collaboration}, ``{Study of the underlying event at
  forward rapidity in pp collisions at $\sqrt{s}$ = 0.9, 2.76, and 7 TeV}'',}
  \textit{ JHEP} \textbf{ 04} (2013) 072,
  \href{http://dx.doi.org/10.1007/JHEP04(2013)072}{\doi{10.1007/JHEP04(2013)072}},
\href{http://www.arXiv.org/abs/1302.2394}{\texttt{arXiv:1302.2394}}.

\bibitem{herwig}
G.~Corcella\hrefCMSnoop {}{ {et~al.}, ``{HERWIG} 6: an event generator for
  hadron emission reactions with interfering gluons (including supersymmetric
  processes)'',} \textit{ JHEP} \textbf{ 01} (2001) 010,
  \href{http://dx.doi.org/10.1088/1126-6708/2001/01/010}{\doi{10.1088/1126-6708/2001/01/010}},
\href{http://www.arXiv.org/abs/hep-ph/0011363}{\texttt{arXiv:hep-ph/0011363}}.

\bibitem{jimmy}
\hrefCMSnoop {}{J.~M. Butterworth, J.~R. Forshaw, and M.~H. Seymour,
  ``Multiparton interactions in photoproduction at {HERA}'',} \textit{ Z. Phys.
  C} \textbf{ 72} (1996) 637,
  \href{http://dx.doi.org/10.1007/s002880050286}{\doi{10.1007/s002880050286}},
\href{http://www.arXiv.org/abs/hep-ph/9601371}{\texttt{arXiv:hep-ph/9601371}}.

\bibitem{cteq6l}
J.~Pumplin\hrefCMSnoop {}{ {et~al.}, ``{New generation of parton distributions
  with uncertainties from global QCD analysis}'',} \textit{ JHEP} \textbf{ 07}
  (2002) 012,
  \href{http://dx.doi.org/10.1088/1126-6708/2002/07/012}{\doi{10.1088/1126-6708/2002/07/012}},
  \href{http://www.arXiv.org/abs/hep-ph/0201195}{\texttt{arXiv:hep-ph/0201195}}.

\bibitem{geant}
\hrefCMSnoop {}{{GEANT4} Collaboration, ``{GEANT4}---a simulation toolkit'',}
  \textit{ Nucl. Instrum. Meth. A} \textbf{ 506} (2003) 250,
\href{http://dx.doi.org/10.1016/S0168-9002(03)01368-8}{\doi{10.1016/S0168-9002(03)01368-8}}.

\bibitem{jq}
\hrefCMSnoop {}{{CMS Collaboration}, ``Determination of jet energy calibration
  and transverse momentum resolution in {CMS}'',} \textit{ J. Instrum.}
  \textbf{ 6} (2011) P11002,
  \href{http://dx.doi.org/10.1088/1748-0221/6/11/P11002}{\doi{10.1088/1748-0221/6/11/P11002}},
  \href{http://www.arXiv.org/abs/1107.4277}{\texttt{arXiv:1107.4277}}.

\bibitem{bscrap}
\hrefCMSnoop {}{{CMS Collaboration}, ``Transverse-momentum and pseudorapidity
  distributions of charged hadrons in pp collisions at $\sqrt{s}$~=~0.9 and
  2.36~{TeV}'',} \textit{ JHEP.} \textbf{ 02} (2010) 041,
  \href{http://dx.doi.org/10.1007/JHEP02(2010)041}{\doi{10.1007/JHEP02(2010)041}},
  \href{http://www.arXiv.org/abs/1002.0621}{\texttt{arXiv:1002.0621}}.

\bibitem{ua5}
\hrefCMSnoop {}{{UA5} Collaboration, ``Multiplicity distributions in different
  pseudorapidity intervals s at a {CMS} energy of {540 GeV}'',} \textit{ Phys.
  Lett. B} \textbf{ 160} (1985) 193,
\href{http://dx.doi.org/10.1016/0370-2693(85)91491-1}{\doi{10.1016/0370-2693(85)91491-1}}.

\bibitem{ua5failstart}
\hrefCMSnoop {}{{UA5} Collaboration, ``Charged particle multiplicity
  distributions at {200 GeV} and {900 GeV} c.m. energy'',} \textit{ Z. Phys. C}
  \textbf{ 43} (1989) 357,
\href{http://dx.doi.org/10.1007/BF01506531}{\doi{10.1007/BF01506531}}.

\bibitem{alice}
\hrefCMSnoop {}{{ALICE Collaboration}, ``{Charged-particle multiplicity
  measurement in proton-proton collisions at $\sqrt{s}=7$ TeV with ALICE at
  LHC}'',} \textit{ Eur. Phys. J. C} \textbf{ 68} (2010) 345,
  \href{http://dx.doi.org/10.1140/epjc/s10052-010-1350-2}{\doi{10.1140/epjc/s10052-010-1350-2}},
\href{http://www.arXiv.org/abs/1004.3514}{\texttt{arXiv:1004.3514}}.

\bibitem{glm}
\hrefCMSnoop {}{E.~Gotsman, E.~Levin, and U.~Maor, ``{Energy dependence of the
  survival probability of large rapidity gaps}'',} \textit{ Phys. Lett. B}
  \textbf{ 438} (1998) 229,
  \href{http://dx.doi.org/10.1016/S0370-2693(98)00972-1}{\doi{10.1016/S0370-2693(98)00972-1}},
\href{http://www.arXiv.org/abs/hep-ph/9804404}{\texttt{arXiv:hep-ph/9804404}}.

\bibitem{Khoze:2013dha}
\hrefCMSnoop {}{V.~A. Khoze, A.~D. Martin, and M.~G. Ryskin, ``{Diffraction at
  the LHC}'',} \textit{ Eur. Phys. J. C} \textbf{ 73} (2013) 2503,
  \href{http://dx.doi.org/10.1140/epjc/s10052-013-2503-x}{\doi{10.1140/epjc/s10052-013-2503-x}},
\href{http://www.arXiv.org/abs/1306.2149}{\texttt{arXiv:1306.2149}}.

\bibitem{Gotsman:2015aga}
\hrefCMSnoop {}{E.~Gotsman, E.~Levin, and U.~Maor, ``{CGC/saturation approach
  for soft interactions at high energy: survival probability of central
  exclusive production}'',} \textit{ Eur. Phys. J. C} \textbf{ 76} (2016) 177,
  \href{http://dx.doi.org/10.1140/epjc/s10052-016-4014-z}{\doi{10.1140/epjc/s10052-016-4014-z}},
\href{http://www.arXiv.org/abs/1510.07249}{\texttt{arXiv:1510.07249}}.

\end{thebibliography}\endgroup

\cleardoublepage \appendix\section{The CMS Collaboration \label{app:collab}}\begin{sloppypar}\hyphenpenalty=5000\widowpenalty=500\clubpenalty=5000\textbf{Yerevan Physics Institute,  Yerevan,  Armenia}\\*[0pt]
A.M.~Sirunyan, A.~Tumasyan
\vskip\cmsinstskip
\textbf{Institut f\"{u}r Hochenergiephysik,  Wien,  Austria}\\*[0pt]
W.~Adam, E.~Asilar, T.~Bergauer, J.~Brandstetter, E.~Brondolin, M.~Dragicevic, J.~Er\"{o}, M.~Flechl, M.~Friedl, R.~Fr\"{u}hwirth\cmsAuthorMark{1}, V.M.~Ghete, C.~Hartl, N.~H\"{o}rmann, J.~Hrubec, M.~Jeitler\cmsAuthorMark{1}, A.~K\"{o}nig, I.~Kr\"{a}tschmer, D.~Liko, T.~Matsushita, I.~Mikulec, D.~Rabady, N.~Rad, B.~Rahbaran, H.~Rohringer, J.~Schieck\cmsAuthorMark{1}, J.~Strauss, W.~Waltenberger, C.-E.~Wulz\cmsAuthorMark{1}
% ...
\vskip\cmsinstskip
% ...
1:~~Also at Vienna University of Technology, Vienna, Austria\\
% ...

\end{sloppypar}
\end{document}